\documentclass[sn-mathphys,Numbered,iicol]{sn-jnl}%
\usepackage{graphics,graphicx}%
\usepackage{multirow,makecell}%
\usepackage{amsmath,amssymb,amsfonts}%
\usepackage{amsthm}%
\usepackage{mathrsfs}%
\usepackage{xcolor}%
\usepackage{textcomp}%
\usepackage{manyfoot}%
\usepackage{booktabs}%
\usepackage{algorithm}%
\usepackage{algorithmicx}%
\usepackage{algpseudocode}%
\usepackage{listings}%
\usepackage{array}
\usepackage{caption2}

\begin{document}

\newcolumntype{M}[1]{>{\centering\arraybackslash}m{#1}}

\title[Article Title]{Exact solutions and Dynamical phase transitions in the Lipkin-Meshkov-Glick model with Dual nonlinear interactions}


\author*[1,2]{\fnm{Yu} \sur{Dongyang}}\email{dyyu21@zjut.edu.cn}

\affil[1]{\orgdiv{Department of Physics}, \orgname{Zhejiang University of Technology}, \orgaddress{\street{No. 288, Liuhe Road}, \city{Hangzhou}, \postcode{310023}, \state{Zhejiang}, \country{China}}}

\affil*[2]{\orgdiv{Zhejiang Provincial Key Laboratory and Collaborative Innovation Center for Quantum Precise Measurement}, \orgname{Zhejiang University of Technology}, \orgaddress{\street{No. 288, Liuhe Road}, \city{Hangzhou}, \postcode{310023}, \state{Zhejiang}, \country{China}}}

\abstract{Lipkin-Meshkov-Glick (LMG) model is paradigmatic to study quantum phase transition in equilibrium or non-equilibrium systems and entanglement dynamics in a variety of disciplines. The generic LMG model usually incorporates two nonlinear interactions. 
While the classical dynamics of the single-nonlinear-inteaction LMG model is well understood through Jacobi elliptic functions, the dual-interaction case remains unexplored due to analytical challenges. Here, by constructing an auxiliary function that maps the dynamics to the complex plane of Jacobi elliptic functions, we derive exact solutions of classical dynamics for the dual-interaction LMG model. Based on the exact solutions, we give the classical dynamical phase diagram of the LMG model with dual nonlinear interactions, and find out a non-logarithmic behavior of dynamical criticality which is absent in case of single nonlinear interaction.
Our results establish a benchmark to analyze the quantum dynamical phase transitions and  many-body entanglement dynamics of finite-size LMG model.}




\maketitle

\section{Introduction}\label{sec1}

Lipkin-Meshkov-Glick (LMG) model in nuclear physics describes $N$ spins interacting with infinite-range interaction~\cite{lipkin1965,meshkov1965,glick1965}, which also can be considered as a special limit of XY Heisenberg model in condensed matter physics~\cite{carrasco2016}. Particularly in recent decades, LMG model has already been engineered in various realistic platforms, i.e., a Bose-Einstein condensate (BEC) in a double well potential~\cite{raghavan1999} or two-component/mode BECs~\cite{zibold2010}, quantum gases in cavity QED~\cite{baumann2010,muniz2020}, thermal Bose gases~\cite{chuAJ2020}, superconducting circuits~\cite{xu2020} and acoustics~\cite{Zhou2024}. 
These experimental researches on LMG model make progress in its potential utilization in quantum information and precise measurement~\cite{pezze2018}. 
Although the exact energy spectra of integrable LMG model has already been obtained~\cite{morita2006,Ribeiro2008} based on Bethe ansatz, 
large number of works~\cite{Unanyan2003,Ribeiro2008,Orus2008,Ma2009,Engelhardt2013,Engelhardt2015,Schliemann2015,lerose2018,Bao2020,Nader2021,Romero2022} are still being devoted into LMG and related models to figure out unexplored exotic phenomena, for instance, chaotic dynamics induced by nonequilibrium quantum fluctuations~\cite{lerose2018}. 

The Hamiltonian of LMG model with dual nonlinear interactions can be written down as follow, 
\begin{equation}
	{\hat H}=h\hat J_x+\frac{1}{N}\big( g_1 {\hat J}_y^2+g_2 {\hat J}_z^2 \big),
	\label{eq:Hamiltonian0}
\end{equation}
where the collective spin operators $\hat J_{\alpha}=\sum_{j=1}^N\frac{1}{2}\sigma_j^\alpha$ ($\alpha=x,y,z$) satisfying the SU$(2)$ algebra, $[\hat J_x,\hat J_y]=i\hat J_z$($\hbar\equiv 1$ is set in this work), $h$ is the magnetic field along $x$ direction, $g_{1,2}$ is the nonlinear interaction among spins align to $y\ (z)$ direction. Considerable attentions are payed to the largest angular momentum sector (the total angular momentum is conserved), specifically  $J=2N$ to which the ground state belongs. In case of implementing LMG model in a setup of two-component/mode BECs~\cite{raghavan1999,viscondi2009,zibold2010,opatrny2015}, 
$\hat J_x=(\hat a^\dagger \hat b+\hat b^\dagger \hat a)/2$, $\hat J_y=(\hat a^\dagger \hat b-\hat b^\dagger \hat a)/(2i)$, $\hat J_z=(\hat a^\dagger \hat a-\hat b^\dagger \hat b)/2$. $\hat a$ and $\hat b$ represents the annihilation operator of Bosonic component (mode), $[\hat a, \hat a^\dagger]=1$ and $[\hat b, \hat b^\dagger]=1$. 
Here it is worthy noting that we focus on the isolated dynamics of this generic LMG model, therefore its energy is also conserved. 
According to the value of nonlinear interactions, two scenarios can be classified, one is $g_1g_2=0$ and the other is $g_1g_2\neq 0$.

The first scenario only involving one nonlinear interaction, i.e., $g_1=0$($g_2\neq0$) or $g_2=0$($g_1\neq0$) whenever Eq.~(\ref{eq:Hamiltonian0}) holds a $Z_2$ symmetry, is widely utilized to generate spin-squeezing~\cite{ma2011} and many-body entanglement~\cite{micheli2003,pezze2018} by dynamics setup, where the one-tangle and concurrence can be estimated by the classical solutions~\cite{vidal2004,lerose2019} that has already been obtained in terms of Jacobi elliptic function (JEF)~\cite{raghavan1999}. 
Additionally, Eq.~(\ref{eq:Hamiltonian0}) in this scenario is also broadly used to study dynamical phase transition (DPT), a universal phenomenon that is triggered by a sudden quantum quench of control parameter~\cite{marino2022} and quantified by nonequilirium order parameters.

At the same time, the second scenario with dual interactions, specifically $g_1\neq 0$ and $g_2\neq 0$, involves much more comprehensive physics than the first one. For example, the emergent supersymmetry~\cite{Unanyan2003,Unanyan2005}, the generation of many-particle entanglement with a large energy gap~\cite{Unanyan2003}, excited-state quantum phase transitions~\cite{Engelhardt2015,Nader2021}, departures from the sub-manifold of coherent states in the Hamiltonian time evolution~\cite{Schliemann2015}, a flurry of works on the second scenario have appeared to date. 
However, the analysis on entanglement dynamics of LMG model heavily relies on the associated analytical solutions in thermodynamic limit, that is, the entanglement dynamics can be checked only in the first scenario analytically because of the lack of analytical solutions of the second scenario~\cite{vzunkovivc2016}. 
Currently, Gaussian or semiclassical methods that can treat interactions non-perturbatively in some limit (i.e. $N\rightarrow \infty$) are the main approaches to deal with DPTs~\cite{das2006,lerose2018}.

%Until now, however, the second scenario involving two nonlinear interactions, specifically $g_1\neq 0$ and $g_2\neq 0$ whenever Eq.~(\ref{eq:Hamiltonian0}) holds a $Z_2\times Z_2$ symmetry, has received little attentions. This lack of focus is largely due to the challenges in implementing this scenario in experimental setups, as it is significantly more difficult than the first scenario~\cite{zibold2010,opatrny2015}. 
%While in terms of theoretical progress, to our best knowledge, the analytical solutions of classical dynamics of generic LMG model Eq.~(\ref{eq:Hamiltonian0}) in the thermodynamic limit are still lack so far~\cite{vzunkovivc2016}, which prevents to investigate the LMG dynamics including its entanglement and DPT. 

Here by constructing an auxiliary fucntion~\cite{salas2022}, we successfully map the classical dynamics of LMG model (Eq.~(\ref{eq:Hamiltonian0})) for arbitrary couplings into the complex plane of JEFs, which provides us a new perspective to investigate semiclassical and quantum dynamics of LMG mdoel.  
Based on this construction, we apply our analytical results to study the corresponding DPTs in the thermodynamic limit after a quench of nonlinear interactions. We give its generic dynamical phase diagram in terms of time-averaged order parameter  and show that its DPTs are fully controlled by  saddle points of isoenergetic surface. Furthermore, we ascertain that the dynamical criticality depends on the choice of time-averaged order parameter,  which is absent in the first scenario. Finally we discuss the detection of DPTs in terms of BECs trapped in a homogeneous toroidal potential.  

This article is arranged as follow. In Section~\ref{sec:II}, we detail the construction procedure of this auxiliary function that is a linear combination of $y$ and $z$ component of macroscopic spin, and briefly discuss the classifications of our analytical solutions (while the full analysis of each regime is left in Appendix~\ref{Appendix:D}).  Then in Section~\ref{sec:III}, we apply our analytical results to discuss the corresponding DPTs after a quench and its detection in a setup of BEC experimentally. In Section~\ref{sec:IV}, we give a conclusion and outlook.

\section{Analytical solutions}\label{sec:II}
Without loss of generality, the strength of magnetic field along $x$ direction $h\equiv 1$ is set, the Heisenberg equations of collective spin operators, 
\begin{align}
	&\partial_t \hat J_x=\frac{g_1-g_2}{N}(\hat J_y \hat J_z+\hat J_z\hat J_y),\nonumber\\
	&\partial_t \hat J_y= -\hat J_z+\frac{g_2}{N}(\hat J_z \hat J_x +\hat J_x\hat J_z),\nonumber\\
	&\partial_t \hat J_z= \hat J_y-\frac{g_1}{N}(\hat J_y \hat J_x +\hat J_x\hat J_y).
\end{align}
As $N$ approaches infinity, the quantum fluctuations in LMG model eventually diminish to zero, which can be understood by intuitively introducing the 'renormalized' operators, $2\hat J_{x,y,z}/N$, 
\begin{eqnarray}
	[\frac{2\hat J_x}{N},\frac{2\hat J_y}{N}]=i\hbar_{\rm eff} \frac{2\hat J_z}{N}
\end{eqnarray}
with an effective Plank's constant $\hbar_{\rm eff}=2/N$ characterizing the quantum fluctuations of finite-size LMG model. Therefore, $\hbar_{\rm eff}\rightarrow 0$ in the thermodynamic limit, and the spin is classical. 
Let $S_{j}=2\langle\Psi_0|\hat J_{j}(t)|\Psi_0\rangle/N$ where the initial state $|\Psi_0\rangle$ is a spin coherent state, as a result, 
\begin{align}
	&\partial_t S_x =(g_1-g_2)S_yS_z,\nonumber\\
	&\partial_t S_y =(-1+g_2S_x)S_z,\nonumber\\
	&\partial_t S_z =(1-g_1S_x)S_y.
	\label{eqn:Sxyz}
\end{align}
constrained by the total angular momentum $S_x^2+S_y^2+S_z^2=1$. Additionally, 
Eqs.~(\ref{eqn:Sxyz}) are also imposed to evolve on the isoenergetic surface with energy $f_0=2\langle\Psi_0|\hat H|\Psi_0\rangle/N$ because we only consider an isolated system. 

Before discussing the generic classical dynamics of Eq.~(\ref{eq:Hamiltonian0}), let us briefly review the classical solutions already known. 
When $g_1=g_2=g$, $S_x$ is also a constant, a $U(1)$ symmetry emerges, 
\begin{equation}
	\partial_t^2 S_{y,z}=-(1-gS_x)^2S_{y,z},
\end{equation}
constrained by $f_0=S_x+\frac{g}{2}(1-S_x^2)$. 
The solution can be written down as follow, $S_x=(1-\sqrt{D})/g$ ($S_x=f_0$ when $g=0$), 
\begin{eqnarray}
	S_{y,z}(t)=\pm A\sin(\sqrt{D}t+\psi),
	\label{eqn:U1}
\end{eqnarray}
with $A=2\sqrt{(\sqrt{D}+f_0g-1)/g^2}$, $D=g^2-2f_0g+1$, the phase $\psi$ is determined by the initial condition.  

In the first scenario $g_1g_2=0$ (here we choose the case $g_1 = 0$ and $g_2\neq 0$), the continuous U(1) symmetry decades to a Z$_2$ one by nonzero $|g_1-g_2|$,  therefore, the dynamics becomes complicated. 
Nonetheless, Eq.~(\ref{eqn:Sxyz}) can be transformed into a first-order nonlinear differential equation, 
\begin{align}
	1-f_0^2=
	(\partial_t S_z)^2+(1-g_2f_0)S_z^2+\frac{g_2^2}{4}S_z^4.
	\label{eqn:Sz}
\end{align}
The left-hand-side of Eq.~(\ref{eqn:Sz}) can be understood as a linear combination between the square of energy and total angular momentum.
While the right hand of Eq.~(\ref{eqn:Sz}) actually means that a classical particle moves in a double well potential, $V(S_z)=\left((1-g_2f_0)S_z^2+g_2^2S_z^4/4\right)/2$. Hence,
the solution of $S_z$ Eq.~(\ref{eqn:Sz}) is identical to one member of JEFs~\cite{raghavan1999,vidal2004,zibold2010} after a simple transformation also seen in next section. 
For the second scenario $g_1g_2\neq 0$, however, the classical dynamics of LMG model becomes even more complex by the dual  nonlinear interactions so that attempts to construct the analytical solutions of Eq.~(\ref{eqn:Sxyz}) in the second scenario rarely success until recently. 

Usually a symmetry analysis on Eq.~(\ref{eqn:Sxyz}) would be useful to figure out the complexity of classical dynamics. 
Suppose that $\{S_x,S_y,S_z\}$ is the solution of Eq.~(\ref{eqn:Sxyz}) at a given parameter set $\{g_1,g_2,f_0\}$ ($f_0$ the mean energy), then $\{-S_x,S_y,S_z\}$ is also the solution at $\{-g_1,-g_2,-f_0\}$. Furthermore, we observe that the solution is invariant under the transformation $S_{y,z}\rightarrow S_{z,y}$, $g_{1,2}\rightarrow g_{2,1}$ and $t\rightarrow -t$ because of time-reversal-like symmetry. Therefore, 
for clarity, \textit{in this section we will only solve Eq.~(\ref{eqn:Sxyz}) in case of $f_0\geq 0$ and i.e., $g_1\geq g_2$ unless specified}. While the solutions in the remained regimes can be obtained accordingly. 

The classical dynamics Eq.~(\ref{eqn:Sxyz}) is greatly dominated by landscape of $f_0$, 
\begin{align}
	f_0=S_x+\frac{1}{2}\big( g_1 S_y^2+g_2 S_z^2 \big).
\end{align}
that is to say, the minimum, saddle-point and the maximum of $f_0$. 
Besides six trivial extremes, a careful analysis on $f_0$ shows that two additional saddle points emerge (for the detail please see Appendix:~\ref{Appendix:C}) when $|g_1|\geq 1$ or $|g_2|\geq 1$.  Importantly, we discover that the position of each of these saddle points depends solely on one interaction parameter, either $g_1$ or $g_2$, rather than both simultaneously. 
This feature makes that the classical temporal evolution of $\langle \hat J_y\rangle$ and $\langle \hat J_z\rangle$ looks like independent.  
As a consequence, it is possible for us to construct an auxiliary function that represents the classical dynamics of Eq.~(\ref{eqn:Sxyz}) by a single variable, as discussed below. 

Firstly, defining an auxiliary function $X(t)=a_1S_y(t)+a_2S_z(t)$ with  $a_1^2=g_1(g_1-g_2)/(g_1^2+g_2^2)$ and $a_2^2=g_2(g_2-g_1)/(g_1^2+g_2^2)$, we find that the classical dynamics of $X$ can be written down in a simple formula, 
\begin{eqnarray}
	\partial_t^2 X+DX+\frac{g_1^2+g_2^2}{2}X^3=0,
	\label{eqn:X-0}
\end{eqnarray}
with an effective frequency $D$,
\begin{equation}
	D=1+g_1g_2-f_0(g_1+g_2).
\end{equation}
Then, Eq.~{(\ref{eqn:X-0})} could be further simplified as a first-order ordinary but nonlinear differential equation constrained by the conservation of angular momentum and energy $f_0$, 
\begin{eqnarray}
	&(\partial_t X)^2+D X^2+\frac{g_1^2+g_2^2}{4}X^4= (1-f_0^2)\frac{(g_1-g_2)^2}{g_1^2+g_2^2},\nonumber\\
	\label{eqn:X-2}
\end{eqnarray}
The first scenario can be considered as a special case of the second scenario because 
when $g_1=0$ and $g_2\neq 0$, $a_1=0$ and $a_2=1$, $X=S_z$. As a result,  Eq.~(\ref{eqn:X-2}) returns to Eq.~(\ref{eqn:Sz}).
When $g_1\neq 0$ and $g_2= 0$, $a_1=1$ and $a_2=0$, $X=S_y$ has a similar formula.  
While for the second scenario, we argue that our generalization to the second scenario lies on a fact that the position of each of emergent saddle points when $|g_1|>1$ and $|g_2|>1$ depends solely on one nonlinear interaction, either $g_1$ or $g_2$, rather both of them. 
When $X\in \mathcal{R}$, i.e. $g_2<0<g_1$, Eq.~(\ref{eqn:X-0}) is a special case of Duffing's equation with no damping and driving force, and considered as an analytic continuation of Duffing's equation into the complex plane when $X\in\mathcal{C}$, i.e., $g_1>g_2>0$ (please see the discussions in Appendix:~\ref{Appendix:A}). Nonetheless, this is quite different from the previous works, where $X$ is defined in real axis~\cite{krech1997} or the cubical term is like $|X|^2X$~\cite{salas2022}. 
In a word, the dynamics of $X$ is identical to a classical particle that is governed by a complex double well potential, $V(X)=(DX^2+(g_1^2+g_2^2)X^4/4)/2$ with $X\in\mathcal{C}$.
We note here that strictly speaking, Eq.~(\ref{eqn:X-0})~(\ref{eqn:X-2}) can not be applied to the case $g_1=g_2$, because $a_1=a_2=0$. 

By introducing 
\begin{equation}
	\tau = \frac{\sqrt{g_1^2+g_2^2}}{2}t, 
	\label{eqn:tau}
\end{equation}
we recast Eq.~(\ref{eqn:X-2}) into a more compact form, 
\begin{eqnarray}
	(\partial_\tau X)^2=P[X]\equiv -(X^4+uX^2+v),
	\label{eqn:X}
\end{eqnarray}
with  
\begin{align}
	&u=\frac{4D}{g_1^2+g_2^2},\nonumber\\
	&v=4(f_0^2-1)\frac{(g_1-g_2)^2}{(g_1^2+g_2^2)^2}. 
\end{align}
Basically, the analytical solutions of Eq.~(\ref{eqn:X}) can be classified into four regimes and expressed as one member of the family of JEFs as summarized in Table.~\ref{tab:table1}. In main text, a brief discussion on the solutions is only given, please go to the Appendix~\ref{Appendix:D} for a comprehensive analysis of each regime. 

\begin{figure}
	\includegraphics[width=0.95\linewidth]{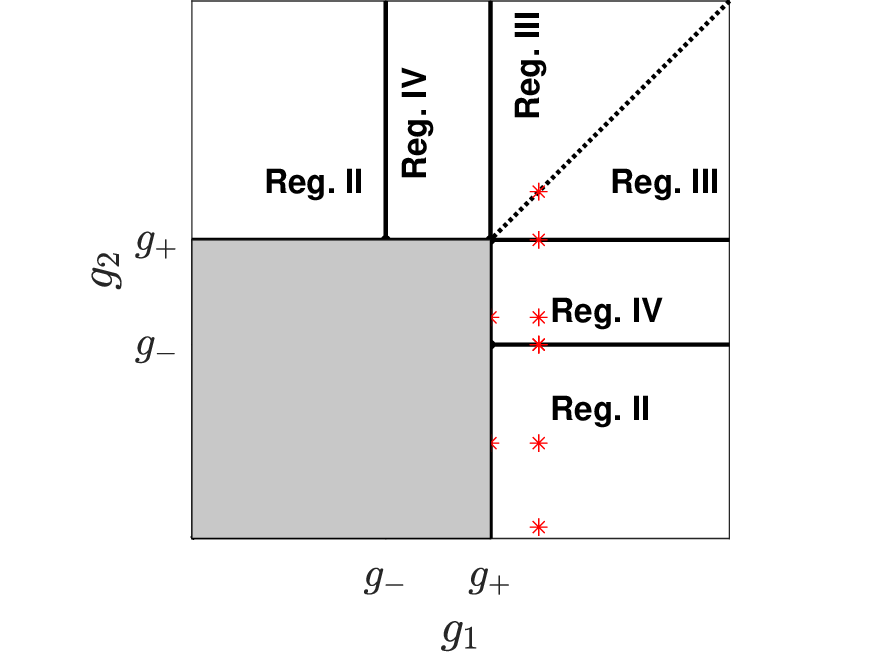}
	\caption{(color online) 
		Solutions of Eq.~(\ref{eqn:X}) can be classified into four regimes in $(g_1, g_2, f_0)$ space. \textbf{Reg.~I}~(Sec.~\ref{sec:solu-I}) always exists in the whole $(g_1, g_2)$ plane when $0\leq f_0\leq 1$. While for $f_0>1$, we take $f_0=1.6$ as an example to explicitly show the regime of \textbf{Reg.~II}~(Sec.~\ref{sec:solu-II}), \textbf{Reg.~III}~(Sec.~\ref{sec:solu-III}) and \textbf{Reg.~IV}~(Sec.~\ref{sec:solu-IV}), the gray regime indicates that no solutions exist there. 
		Typical points (red asterisks) are chosen as examples to demonstrate the classical dynamics and especially show the consistency between  analytical (Eq.~(\ref{eqn:X})) and numerical (Eq.~(\ref{eqn:Sxyz})) solutions, see Fig.~(\ref{fig:example11}, \ref{fig:example121}, \ref{fig:example122}, \ref{fig:example212}). Here the $g_1=g_2$ line is particularly plotted (dotted line) as a guideline.}
	\label{fig:pd}
\end{figure}

Our classification is based on  the sign of $v$, $u$ and the discriminant $\Delta$ of $P[X]=0$,  
\begin{equation}
	\Delta=\frac{4R_1R_2}{(g_1^2+g_2^2)^2},
\end{equation} 
with
\begin{eqnarray}
	R_{1,2}=g_{1,2}^2-2f_0g_{1,2}+1.
\end{eqnarray}
In fact, the critical points set by $\Delta=0$ correspond to the minimum/maximum/saddle points of $f_0$ in parameter space  (Appendix~\ref{Appendix:C}), or equivalently, the critical values of $g_{1,2}$ are given as $g_\pm(f_0)$ for a fixed value of $f_0$, 
\begin{eqnarray}
	g_\pm(f_0)=f_0\pm\sqrt{f_0^2-1}\ \text{as}\ |f_0|\geq 1.
\end{eqnarray}
For clarity, we take $f_0=1.6$ as an example to display the \textbf{Reg.~II}~(Sec.~\ref{sec:solu-II}), \textbf{Reg.~III}~(Sec.~\ref{sec:solu-III}), and \textbf{Reg.~IV}~(Sec.~\ref{sec:solu-IV}) in Fig.~\ref{fig:pd} (where we do not explicitly depict the \textbf{Reg.~I}~(Sec.~\ref{sec:solu-I}), because \textbf{Reg.~I} covers the entire $g_1-g_2$ plane when $0\leq f_0\leq 1$). 
In each regime, Eq.~(\ref{eqn:X}) can be ultimately transformed into the differential equation of a member within the set of JEFs as shown in Eq.~(\ref{eqn:solution11}), (\ref{eqn:solution121}), (\ref{eqn:solution122}), and (\ref{eqn:solution212}).
To fully determine $X(t)$, however, two initial conditions of $X$ are further needed,  $X(0)=a_1S_y(0)+a_2S_z(0)$ and its first order derivative with respect to time $X^\prime(0)=a_1S_y^\prime(0)+a_2S_z^\prime(0)$ (given by Eq.~(\ref{eqn:Sxyz})).
Accordingly, the analytical solutions of Eq.~(\ref{eqn:X}), $S_{x,y,z}(t)$, can be explicitly constructed according to the sign of $a_{1,2}$ (see Appendix~\ref{Appendix:B}) and compared with the numerical ones of Eq.~(\ref{eqn:Sxyz}) which shows an excellent consistency with each other (It is worthy noting that the high-precision numerical solutions of Eq.~(\ref{eqn:Sxyz}) can be easily obtained in a laptop computer with few cores). 

In next section, we use these analytical solutions to demonstrate the DPT features of LMG model in presence of dual nonlinear interactions. 

\section{Dynamical phase transitions}\label{sec:III}
LMG model is widely used to investigate DPT after a quench of controlling parameters~\cite{solinas2008,marino2022}. However, due to the complexity of experimental setups and challenges of theoretical approaches, these discussions are much more focused on the first scenario rather than the second one, both of which could be engineered in various setups~\cite{Baksic2014,Soriente2018} also as discussed in the Introduction (Sec.~\ref{sec1}). In this section, we apply our analytical results to uncover the properties of DPT in the thermodynamic limit especially for the second scenario and discuss the detection of DPTs in the setup of BECs in a toroidal trap~\cite{opatrny2015}. 

\begin{figure}
	\includegraphics[width=0.95\linewidth]{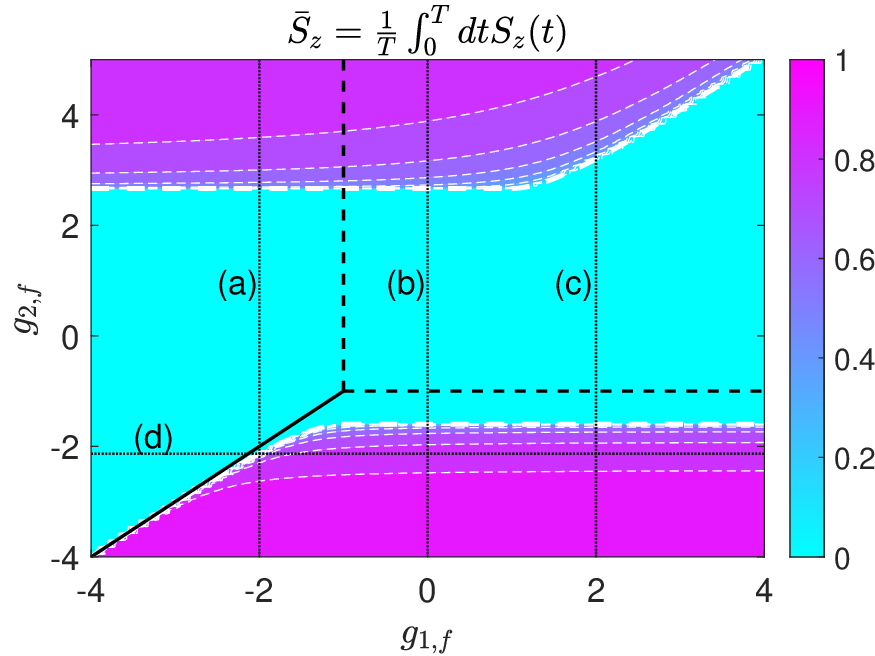}
	\caption{(color online) Dynamical phase diagram of the generic LMG model in terms of time-averaged order parameter $\bar S_z=\int_0^T dt S_z(t)/T$ in the  thermodynamic limit when $g_{2,i}=-4<g_{1,i}\leq-1$. The boundaries of DPT are marked as white thick dashed lines. As a comparison, we also depict the corresponding zero-temperature phase diagram of equilibrium system (black thick solid (first order) and dashed (second order) lines).}
	\label{fig:DPD1}
\end{figure}

Fig.~\ref{fig:DPD1} demonstrates the typical dynamical phase diagram in terms of time-averaged order parameter $\bar S_z=\int_0^T dt S_z(t)/T$ after quenching the nonlinear interactions $(g_1,g_2)$ from the initial values $(g_{1,i},g_{2,i})$ ($g_{1,i}>g_{2,i}$ is imposed) to the final ones $(g_{1,f},g_{2,f})$. Here we take $g_{2,i}=-4$ as an example.  
The initial state is the ferromagnetic ground state, breaking a $Z_2$ symmetry, with $S_{z}(0)=\sqrt{1-g_{2,i}^{-2}}>0$ and $S_y(0)=0$ when $g_{1,i}>g_{2,i}$ and $g_{2,i}\leq -1$. 
It becomes clear that the phase diagram of DPT (marked by the white thick dashed lines in Fig.~\ref{fig:DPD1}) is inconsistent with the one of its equilibrium counterpart (by the black thick dashed (second order)/solid (first order) lines in Fig.~\ref{fig:DPD1}). 
In detail, when $g_{1,f}\geq-1$, the lower boundary of DPT $g_{2,f}^-$ is obtained by  $f_0(g_{1,f},g_{2,f},\theta,\phi)=-1$ (the isoenergetic surface of $f_0$ can be parameterized in terms of the polar and azimuthal angle $(\theta,\phi)$ of Bloch sphere), 
\begin{eqnarray}
	g_{2,f}^- = \frac{-2g_{2,i}}{g_{2,i}-1}.
	\label{eqn:dpd1-1}
\end{eqnarray}
And when $g_{1,f}\leq 1$ the upper boundary $g_{2,f}^+$ is determined by $f_0(g_{1,f},g_{2,f},\theta,\phi)=1$, 
\begin{eqnarray}
	g_{2,f}^+ = \frac{2g_{2,i}}{g_{2,i}+1}.
	\label{eqn:dpd1-2}
\end{eqnarray}
While the lower boundary as $g_{1,f}<-1$ and the upper one as $g_{1,f}>1$ is controlled by $f_0(g_{1,f},g_{2,f},\theta,\phi)=\frac{1}{2}(g_{1,f}+1/g_{1,f})$ and 
written in a unified formula,
\begin{eqnarray}
	&\frac{g_{2,i}(g_{2,i}+g_{1,f}(g_{1,f}g_{2,i}-2))}{g_{1,f}(g_{2,i}^2-1)}=
	\left\{
	\begin{array}{cc}
		g_{2,f}^-, &\ g_{1,f}<-1\\
		g_{2,f}^+, &\ g_{1,f}>1\\
	\end{array}.\right.\nonumber\\
	\label{eqn:dpd1-3}
\end{eqnarray}
As expected, the critical lines of DPT in the thermodynamic limit, i.e., $g_{2,f}^\pm$, can be uniformly understood as the conditions under which the post-quenched energy $f_0(g_{1,f},g_{2,f},\theta,\phi)$ intersects one of the saddle points, namely $f_0=\pm 1$ or $f_0=\frac{1}{2}(g_{1,f}+1/g_{1,f})$. 

\begin{figure}
	\includegraphics[width=0.95\linewidth]{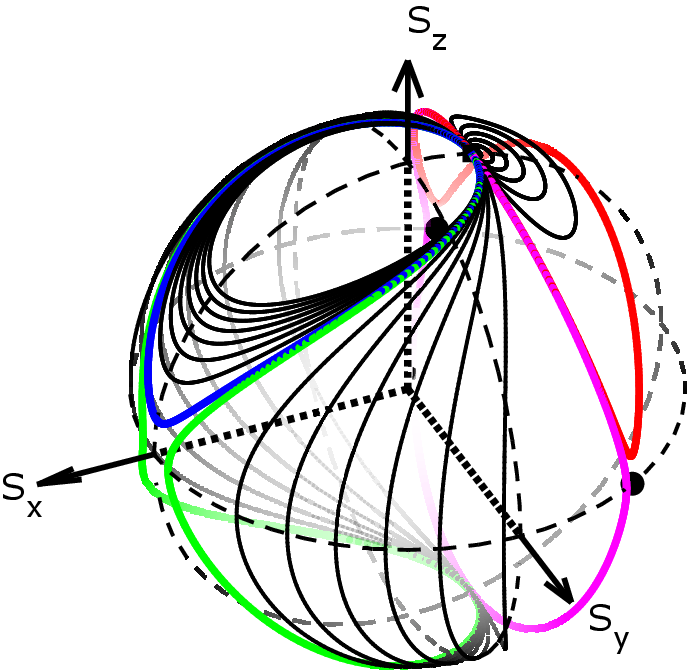}
	\caption{(color online) Quench dynamics of Eq.~(\ref{eqn:Sxyz}) in the thermodynamic limit is depicted on Bloch sphere after a quench $g_{2,f}$
	as  $g_{1,f}=-2$ (labeled as (a) as the black thin dashed line in Fig.~\ref{fig:DPD1}).  
		Here four special trajectories are chosen to emphasize the two DPT transitions, the solid red, magenta, green and blue line represents the relevant classical trajectory with $g_{2,f}=g_{2,f}^--\epsilon$, $g_{2,f}^-+\epsilon$, $g_{2,f}^+-\epsilon$ and $g_{2,f}^++\epsilon$, respectively. Here $\epsilon$ is a positive infinitesimal constant. }
	\label{fig:DPT1-BlochSphere}
\end{figure}

Let us take a specific example to clearly demonstrate the evolution of classical trajectory of LMG model with dual nonlinear interactions in Fig.~\ref{fig:DPT1-BlochSphere} (labeled as (a) in Fig.~\ref{fig:DPD1}). 
When $g_{2,f}<g_{2,f}^-$, the dynamics is largely determined by the global minimum $f_0=(g_{2,f}+1/g_{2,f})$ and trapped in the northern hemisphere of Bloch sphere with finite time-averaged order parameter $\bar S_z$, which is a reminiscent of mesoscopic self-trapping (MST) mode with oscillating phase. 
However, as $g_{2,f}\rightarrow g_{2,f}^-$, the classical trajectory gradually approaches to the one (marked as red solid line in Fig.~\ref{fig:DPT1-BlochSphere}) close to the saddle points $f_0=\frac{1}{2}(g_{1,f}+1/g_{1,f})$ (marked as black solid circles in Fig.~\ref{fig:DPT1-BlochSphere}) and sharply jumps to the one with vanished $\bar S_z$ (magenta solid line in Fig.~\ref{fig:DPT1-BlochSphere}) as $g_{2,f}$ crossing the critical point $g_{2,f}^-$. Therefore, a DPT occurs at $g_{2,f}=g_{2,f}^-$ and the $Z_2$ symmetry restores again. 
Here it is worthy noting that the two saddle points (black solid circles) finally merge with each other at $(\theta,\phi)=(\pi/2,\pi)$ as $g_{1,f}\rightarrow 0$. 
Furthermore, when $g_{2,f}^-<g_{2,f}\leq g_{2,f}^+$, the dynamics holds this $Z_2$ symmetry with $\bar S_z=0$ until its trajectory intersects the second saddle point $(\theta,\phi)=(\pi/2,0)$ ($f_0=1$). 
Finally, another DPT occurs and breaks its corresponding $Z_2$ symmetry again as $g_{2,f}$ crossing $g_{2,f}^+$ (exemplified by the solid green and blue lines in Fig.~\ref{fig:DPT1-BlochSphere}). However, these emerged MST modes are different from the ones when  $g_{2,f}<g_{2,f}^-$, due to the nature of running phase. 

\begin{figure}
	\includegraphics[width=0.95\linewidth]{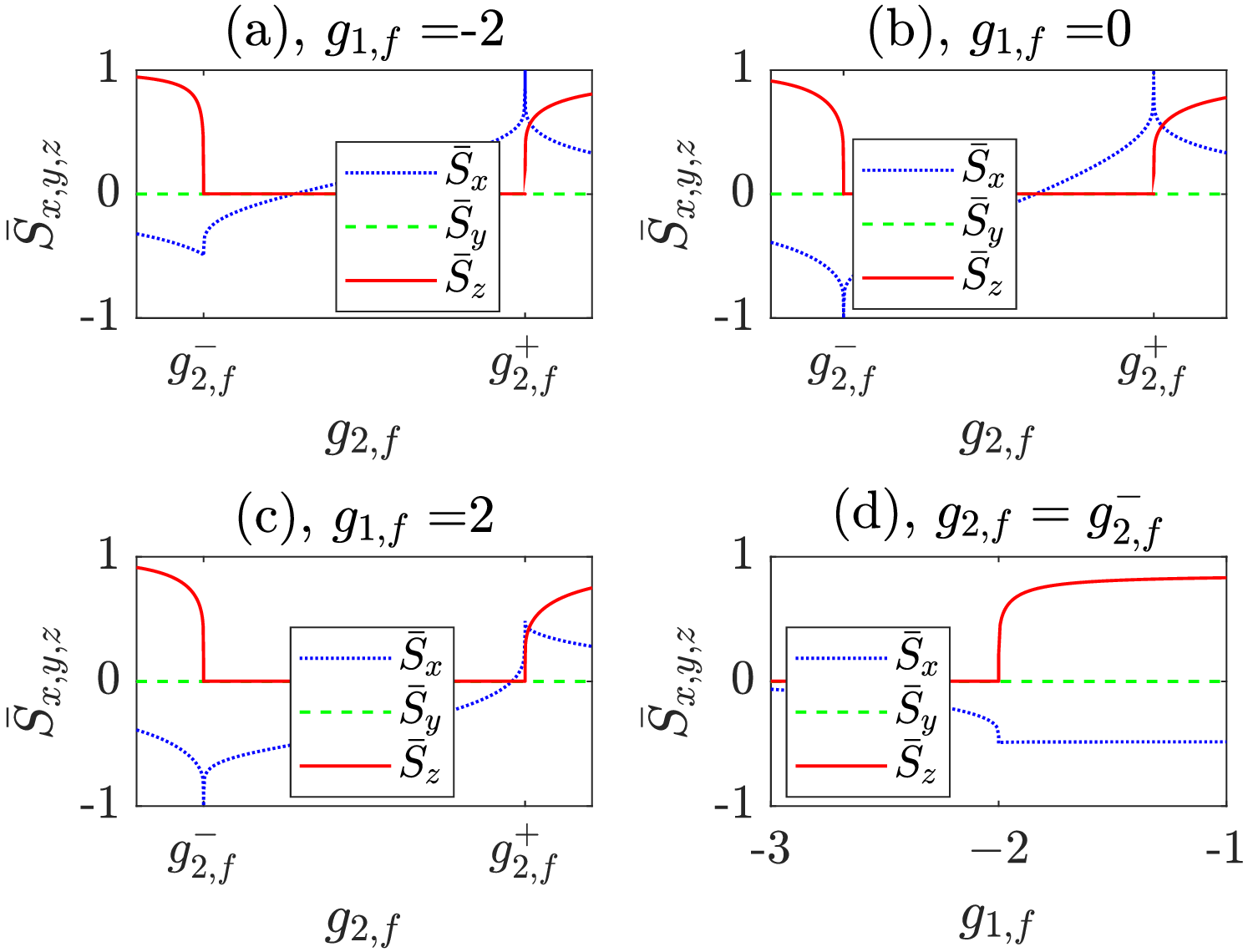}
	\caption{(color online) DPT as quenching control parameters from i.e., $(g_{1,i},g_{2,i}=-4)$ (take $g_{1,i}>g_{2,i}$ to ensure the initial state is the ferromagnetic ground state) to the final parameters, $(g_{1,f},g_{2,f})$. Here we choose $g_{1,f}=-2$ (a), $0$ (b), $2$ (c) while  varying $g_{2,f}$, and $g_{2,f}=g_{2,f}^-$ (d) as varying $g_{1,f}$ to demonstrate the evolution of time-averaged order parameters $\bar S_{x,y,z}$. Here $\bar S_y$ is always zero.}
	\label{fig:DPT1}
\end{figure}

Distinct from its equilibrium counterpart, the time-averaged order parameter of DPT usually displays a logarithmic singularity~\cite{BLi2019} around dynamical critical point (i.e., $g_{2,f}^\pm$) as depicted in Fig.~\ref{fig:DPT1} as quenching $g_2$ at a fixed $g_1$. 
However, we find out the criticality around $g_{2,f}^\pm$ looks like depends on the choice of order parameter and value of $g_{1,2}$, especially, the singularity of ${\bar S}_x$ around critical point shows non-logarithmic behaviors in case of a non-zero value of $g_1$ in the second scenario. This non-logarithmic phenomena requires further investigations.  

At last, 
we briefly discuss the corresponding detection of DPTs in terms of BEC trapped in a toroidal trap where both scenarios of LMG model (Eq.~\ref{eq:Hamiltonian0}) can be engineered in a realistic setup, following the Ref.~\cite{opatrny2015}. To engineer the LMG model Eq.~(\ref{eq:Hamiltonian0}), 
two counter-propagating rotational modes of a homogeneous ring trap are initially populated while the remained modes are gaped away because of the large kinetic energy gap. And the interaction between the two Bosonic modes, imposed by the mechanism of an optical Feshbach resonance, varies  along the azimuth angle, specifically, $g(\phi,t)\propto g_s+g_d\cos(4q(\omega t+\phi))$. Here $g_{s,d}$ represents the strength of static and modulated interaction, both of which can be finely tuned. $q$ is an integer number, $\omega$ is the modulation frequency. 
Then Eq.~(\ref{eq:Hamiltonian0}) is finally obtained under a spin rotation, i.e., $\hat J_{x}\rightarrow \hat J_{y}$, $\hat J_{y}\rightarrow\hat J_z$ and $\hat J_{y}\rightarrow\hat J_z$, with the effective Hamiltonian $\hat H_{\rm eff}=2q\omega\hat J_z+(\frac{\chi_1}{2}+\chi_0)\hat J_x^2+(-\frac{\chi_1}{2}+\chi_0)\hat J_y^2$, where $\hat J_{x,y,z}$ shares the same definition as in the Section~\ref{sec1}, $\hat a\equiv\hat a_{q} e^{-iq\omega t}$ and $\hat b\equiv\hat a_{-q} e^{iq\omega t}$.   
Based on the advanced time-of-flight technique, 
the aforementioned DPTs can be well extracted by measuring its azimuth density $n(\phi,t)=\langle\Psi_0|\hat \psi^\dagger(\phi,t)\hat \psi(\phi,t)|\Psi_0\rangle \propto 1+S_y(t)\cos(2q\omega t+\phi)-S_z(t)\sin(2q\omega t+\phi)$ with $\hat\psi(\phi,t)\propto\hat a_q e^{iq\phi}+\hat a_{-q}e^{-iq\phi}$.

\section{Conclusions and Outlook}\label{sec:IV}
In this work, by constructing an auxiliary function we map the classical dynamics of LMG model with dual nonlinear interactions into the complex plane of Jacobi elliptic functions in a generic framework, and finally obtain the exact solutions of classical dynamics of LMG model in the thermodynamic limit. Based on these analytical results, we further give the dynamical phase diagram in terms of time-averaged order parameter after quenching one of dual nonlinear interactions, and analytically discuss the properties of dynamical phase transitions triggered by the saddle points of isoenergetic surface. Particularly in the second scenario ($g_{1,f}g_{2,f}\neq 0$), we observe a non-logarithmic behavior of dynamical criticality which depends on the choice of time-averaged order parameter and the strength of interaction.

A direct application of our results would be   to evaluate the departures from sub-manifold of spin-coherent states precisely, according to the theorem given by Schliemann~\cite{Schliemann2015}. 
Our analytical solutions in the thermodynamic limit would become a generic benchmark~\cite{Huang2018} on quantum dynamical phase transitions and many-body entanglement dynamics of finite-size LMG model with dual interactions, highlighting the generation of strong many-body entanglement in realistic experiment setup. 
The extension of our results to the presence of dissipation or decoherence would be left as a future work~\cite{stitely2022}.

\backmatter

\bmhead{Acknowledgments}

The author thanks Dr. Siqi Ren for stimulating discussions, and also thanks Prof. Chao Gao for helpful suggestions on organization and English grammar of our manuscript. 
This work was supported by the starting fund of Zhejiang University of Technology. 

\bmhead{Data Availability Statement} This manuscript has no associated data or the data will not be deposited. [Authors comment: The datasets generated and/or analyzed during the current study are available from the corresponding author on reasonable request.]

\begin{appendices}
	
\section{}\label{Appendix:D}
As mentioned in main text, four distinct regimes are classified in the $(g_1,g_2,f_0)$ space, according to the sign of $u$, $v$ and $\Delta$. The brief summation of exact solutions of Eq.~\ref{eqn:Sxyz} are listed in Table~\ref{tab:table1} while the details of each regime are fully discussed in the following subsections. 
\begin{table*}[t]
	\setcaptionwidth{1\linewidth} 
	\centering
	\normalsize
	\begin{tabular}{|>{\centering}m{3.2cm}|>{\centering}m{2cm}|m{8.4cm}|} 
		\hline
		& Range of $f_0$, $g_{1,2}$ & $X(t)$  \\ 
		\hline
		\raisebox{-0.2cm}{\textbf{Reg.~I} (Sec.~\ref{sec:solu-I})} & \raisebox{-0.2cm}{$0\leq f_0\leq 1$}  &   
		$\begin{aligned} X(t)=\frac{w_-w_+}{\sqrt{w_-^2+w_+^2}}{\rm sd}(\xi \sqrt{w_-^2+w_+^2}\tau+p_0,m) \end{aligned}$ (Eq.~\ref{eqn:solution11})  \\ 
		\hline
		\raisebox{-0.3cm}{\textbf{Reg.~II} (Sec.~\ref{sec:solu-II})} & $\begin{aligned}&f_0\geq 1,\nonumber\\ &g_2\leq g_-,\nonumber\\  &g_+\leq g_1.\end{aligned}$  &  \raisebox{-0.3cm}{$\begin{aligned} X(t)=w_-\cdot{\rm nd}(\xi w_+ \tau + p_0,m) \end{aligned}$ (Eq.~\ref{eqn:solution121})}  \\ 
		\hline
		\raisebox{-0.3cm}{\textbf{Reg.~III} (Sec.~\ref{sec:solu-III})} & \raisebox{-0.3cm}{$\begin{aligned} &f_0\geq 1,\nonumber\\ &g_+\leq g_2\leq g_1.\end{aligned}$} &  \raisebox{-0.3cm}{$\begin{aligned} X(t)=iw_-\cdot {\rm sn}(\xi w_+\tau+p_0,m) \end{aligned}$ (Eq.~\ref{eqn:solution122})}\\ 
		\hline
		\raisebox{-0.3cm}{\textbf{Reg.~IV} (Sec.~\ref{sec:solu-IV})} & \raisebox{-0.3cm}{$\begin{aligned}&f_0\geq 1,\nonumber\\ &g_-\leq g_2\leq g_+.\end{aligned}$} & 
		$\begin{aligned} &X(t)=v^{\frac{1}{4}}\frac{1+Y(t)}{1-Y(t)}\nonumber\\  &Y(t)=i\sqrt{\frac{w_-}{w_+}}\cdot {\rm sn}(2\xi v^{\frac{1}{4}}b_+\tau+p_0,m) \end{aligned}$ (Eq.~\ref{eqn:solution212})\\
		\hline
	\end{tabular}\
	\caption{Four distinct regimes (Reg.~I-IV) of Eq.~(\ref{eqn:X}) are identified based on the signs of $v$, $u$ and $\Delta$, with clear parameter ranges for each regime detailed (the 2nd column). The indeterminate parameters of $X(t)$ in each row should be addressed in the corresponding subsection of Appendix.~\ref{Appendix:D}.}.
	\label{tab:table1}
\end{table*}

\subsection{\textbf{Reg.~I}: $0\leq f_0\leq 1$}\label{sec:solu-I}
The solution of Eq.~(\ref{eqn:X}) is unique when $0\leq f_0\leq 1$. 
By introducing $Y=X\sqrt{w_-^2+w_+^2}/(w_-w_+)$,  $\tau^\prime = \sqrt{w_-^2+w_+^2}\tau$~(Eq.~(\ref{eqn:tau})), and $w_\pm=\sqrt{\sqrt{\Delta}\pm u/2}$, Eq.~(\ref{eqn:X}) is recast as follow (see Appendix~\ref{Appendix:B}),
\begin{equation}
	(\partial_{\tau^\prime} Y)^2=(1-m^\prime Y^2)(1+m Y^2), 
	\label{eqn:Y1}
\end{equation}
with Jacobi elliptic modulus (JEM)
$m=w_-^2/(w_-^2+w_+^2)$. Actually,  
Eq.~(\ref{eqn:Y1}) is identical to the differential equation of JEF ${\rm sd}(z,m)$~\cite{armitage2006}. 
As a consequence, the solution of Eq.~(\ref{eqn:X}) is obtained,
\begin{equation}
	X(t)=\frac{w_-w_+}{\sqrt{w_-^2+w_+^2}}{\rm sd}(\xi \sqrt{w_-^2+w_+^2}\tau+p_0,m),
	\label{eqn:solution11}
\end{equation}
here the parameters $p_0$ and $\xi=\pm 1$ are determined by the initial values $X(t=0)$,
\begin{eqnarray}
	X(0)&=&w_-w_+/\sqrt{w_-^2+w_+^2}\cdot{\rm sd}(p_0,m),\\
	X^\prime(0)&\propto& \xi\cdot {\rm cd}(p_0,m)\cdot {\rm nd}(p_0,m). 
\end{eqnarray}
Because both cd$(z,m)$ and nd$(z,m)$ are even functions of $z$, therefore, 
the temporal evolution frequency $\Omega$ of periodical solution Eq.~(\ref{eqn:X}), 
\begin{eqnarray}
	\Omega=\frac{\pi}{2\sqrt{2}K}(R_1R_2)^{\frac{1}{4}},
\end{eqnarray}
with the complete elliptic integral of the first kind
\begin{eqnarray}
	K(m)=\int_0^{\pi/2}d\varphi(1-m\sin^2(\varphi))^{-1/2}. 
\end{eqnarray}
The definition of $K(m)$ is the same throughout this work. 

\begin{figure}
	\includegraphics[width=0.95\linewidth]{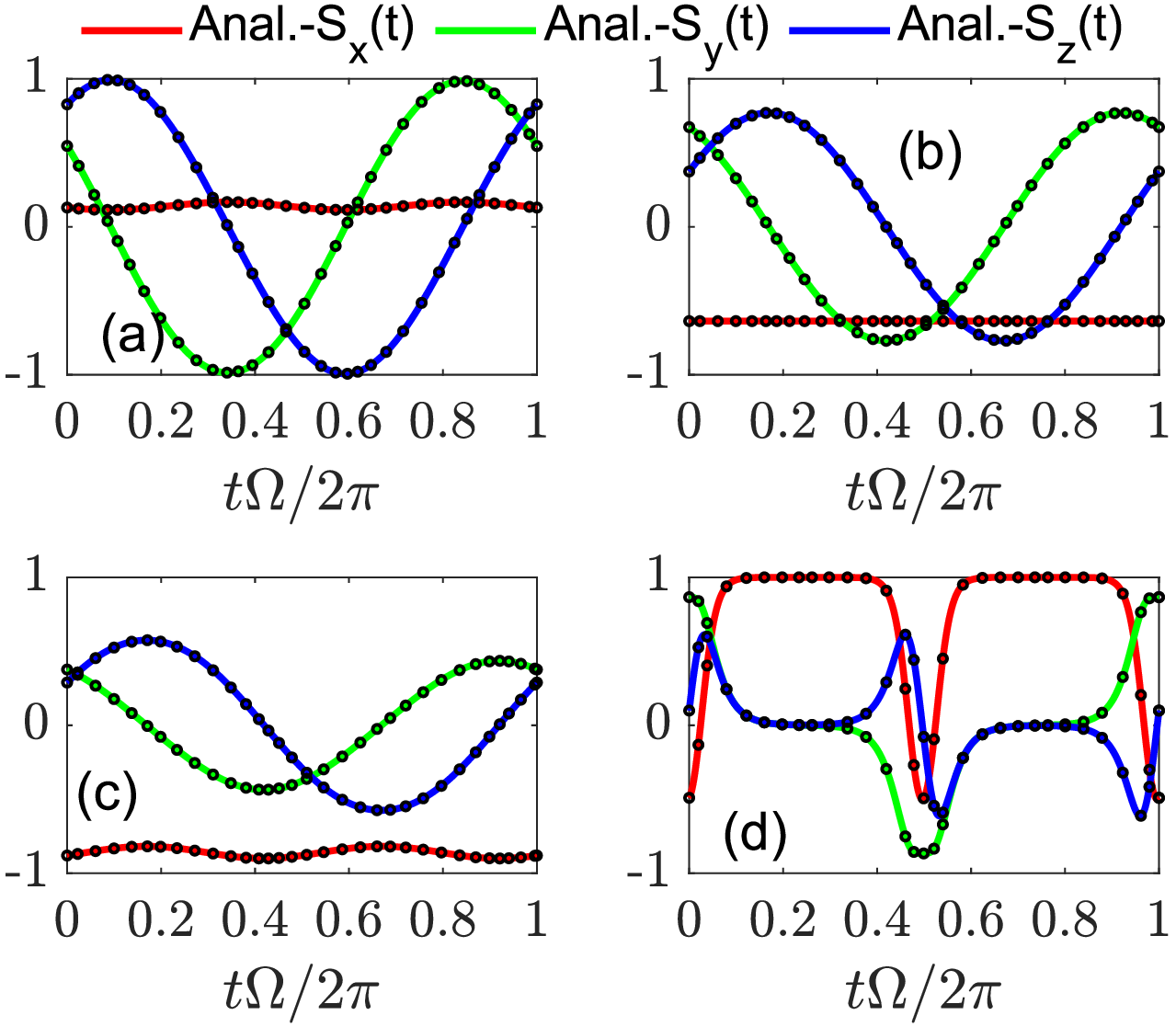}
	\caption{(color online) 
		Comparisons between the analytical solutions Eq.~(\ref{eqn:solution11}) and the numerical ones Eq.~(\ref{eqn:Sxyz}). The solid red, green and blue line corresponds to the $S_x(t)$, $S_y(t)$ and $S_z(t)$, respectively. While the black circles are the corresponding results obtained by numerically solving Eq.~(\ref{eqn:Sxyz}).  
		For subplot (a-d), (a) $g_1=0.5$, $g_2=0.6$, and  $f_0=0.38$; (b), $g_1=4$, $g_2=4-\epsilon$, and  $f_0=0.59$; (c), $g_1=20$, $g_2=11$, and $f_0=1-\epsilon$; (d), $g_1=4$, $g_2=-2$ , and  $f_0=1-\epsilon$. Here $\epsilon$ is an infinitesimal positive real number.}
	\label{fig:example11}
\end{figure}

When $|g_1|\ll 1$ and $|g_2|\ll 1$, the cubic term $X^3$ is negligible.  
Hence, a Rabi oscillation is observed because of the dominated linear term~(\ref{eqn:X}), see Fig.~(\ref{fig:example11})(a).
In general, the temporal evolution of $X$ or $S_{x,y,z}$ depends significantly on the JEM,
\begin{eqnarray}
	m=\frac{1}{2}\left(1-{\rm Sgn}(D)\sqrt{1-\frac{(g_1-g_2)^2(1-f_0^2)}{R_1R_2}}\right),\nonumber\\
\end{eqnarray}
here Sgn$(\cdot)$ means a sign function. 
In case that $g_1$ approaches to $g_2$ ($g_1\rightarrow g$ and $g_2\rightarrow g$, the JEM $m\rightarrow 0$), the dynamics of $S_{x,y,z}$  oscillates sinusoidally with frequency $\Omega\rightarrow \sqrt{g^2-2f_0g+1}$, consistent with the results of Eq.~(\ref{eqn:U1}), see Fig.~(\ref{fig:example11})(b). 

While as $f_0\rightarrow 1$, a simple expression of JEM is obtained, 
\begin{eqnarray}
	m\Big|_{f_0\rightarrow 1} = \frac{1}{2}\left(1-{\rm Sgn}(1-g_1)\cdot {\rm Sgn}(1-g_2)\right),\nonumber\\
\end{eqnarray}
In case $g_1>1$ and $g_2>1$ (or $g_1<1$ and $g_2<1$), the JEM $m\rightarrow 0$ when $f_0\rightarrow 1$, which means an exact Rabi oscillation with  $\Omega\rightarrow \sqrt{(g_1-1)(g_2-1)}$ even though the nonlinear interactions, i.e. $g_1$ and $g_2$, may be very strong (see Fig.~\ref{fig:example11}(c)). 
While $g_1>1$ and $g_2<1$ (or $g_1<1$ and $g_2>1$), $m\rightarrow 1$, $X(t)\propto \sinh(\sqrt{w_-^2+w_+^2}\tau+p_0)$. Therefore, around the initial time, the dynamics of $S_{x,y,z}$, no matter what the strength of $|g_1|$ and $|g_2|$ is, undergoes an exponential growth and saturates around the saddle point of $f_0$ ($(S_x,S_y,S_z)\approx(1,0,0)$) with vanishing frequency $\Omega\rightarrow 0$, as depicted in Fig.~(\ref{fig:example11}(d)).

\subsection{$f_0\geq 1$}
For a fixed $f_0\geq 1$, 
although $v\geq 0$ always holds on (see Table.~\ref{tab:table1}), the solutions of Eq.~(\ref{eqn:X}) still depend on the signs of $\Delta$ and $u$ 
which will change as varying $g_2$, see Table~\ref{tab:table1} and Fig.~\ref{fig:pd}. In general, three regimes are classified further: \textbf{Reg.~II}~(Sec.~\ref{sec:solu-II}) where $g_2\leq g_-(f_0)$, $\Delta\geq 0$, and $u\leq 0$; \textbf{Reg.~III}~(Sec.~\ref{sec:solu-III}) where $g_2\geq g_+(f_0)$, $\Delta\geq 0$ and $u\geq 0$; \textbf{Reg.~IV}~(Sec.~\ref{sec:solu-IV}) where $g_-(f_0)< g_2< g_+(f_0)$, $\Delta\leq 0$. Next we will discuss them individually.

\subsubsection{\textbf{Reg.~II}: $g_2\leq g_-(f_0)$ and $g_+(f_0)\leq g_1$}\label{sec:solu-II}
In this case, $\Delta>0$, $u<0$, and $\sqrt{u^2/4-v}<|u|/2=-u/2$. 
By defining $Y=X/w_-$, $\tau^\prime= w_+\tau$~(Eq.~(\ref{eqn:tau})),, $w_\pm = \sqrt{-u/2\pm \sqrt{\Delta}}$ and the corresponding JEM $m=1-w_-^2/w_+^2$,
\begin{eqnarray}
	m=\frac{2\sqrt{R_1R_2}}{\sqrt{R_1R_2}-D},
\end{eqnarray}
we find that Eq.~(\ref{eqn:X}) can be rewritten down as (Appendix~(\ref{Appendix:A}))  
\begin{equation}
	(\partial_{\tau^\prime} Y)^2=(Y^2-1)(1-(1-m) Y^2), 
	\label{eqn:Y2}
\end{equation}
which is identical to the differential equation of nd$(z,m)$~\cite{armitage2006}. 
Therefore, the analytical solutions of Eq.~(\ref{eqn:X}) 
\begin{eqnarray}
	X(t)=w_-\cdot{\rm nd}(\xi w_+ \tau + p_0,m),
	\label{eqn:solution121}
\end{eqnarray}
here $p_0$ and $\xi=\pm 1$ satisfying 
\begin{eqnarray}
	X(0)&=&w_-\cdot{\rm nd}(p_0,m)\\
	X^\prime(0)&\propto& \xi \frac{{\rm cn}(p_0,m)\cdot {\rm sn}(p_0,m)}{{\rm dn}^2(p_0,m)}. 
\end{eqnarray}
And the frequency $\Omega$, 
\begin{equation}
	\Omega=\frac{\pi}{2\sqrt{2}K}\sqrt{\sqrt{R_1R_2}+D}.
\end{equation}

\begin{figure}
	\includegraphics[width=0.95\linewidth]{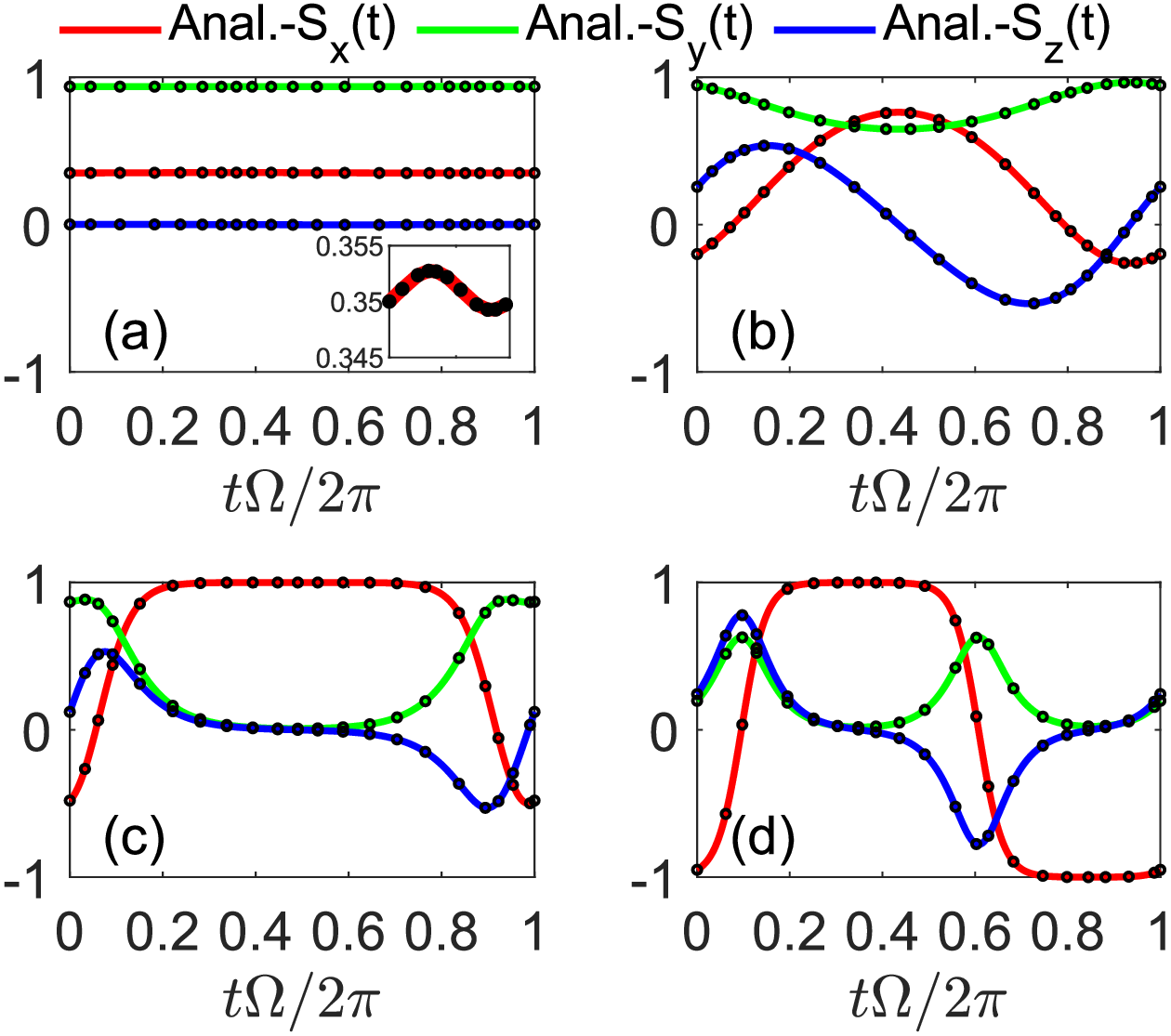}
	\caption{(color online) 
		Comparisons between the analytical solutions Eq.~(\ref{eqn:solution121}) and the numerical ones Eq.~(\ref{eqn:Sxyz}). The solid red, green and blue line corresponds to the $S_x(t)$, $S_y(t)$ and $S_z(t)$, respectively. While the black circles are the corresponding results obtained by numerically solving Eq.~(\ref{eqn:Sxyz}). For subplot (a-d), 
		(a), $g_1=g_+(f_0)+\epsilon$, $g_2=-2$, and $f_0=1.6$; (b), $g_1=4$, $g_2=g_-(f_0)+\epsilon$, and  $f_0=1.6$; (c), $g_1=4$, $g_2=-4$, and $f_0=1+\epsilon$; 
		(d) $g_1=20000$, $g_2=-13000$, and $f_0=4.5$, here $\epsilon$ is an infinitesimal positive real number. An inset in Fig.~\ref{fig:example121}(a) shows the slightly shaking of $S_x(t)$ at a closer scale.}
	\label{fig:example121}
\end{figure}

Let us examine what feature of Eq.~(\ref{eqn:X}) is when $g_1\rightarrow g_+(f_0)$ or $g_2\rightarrow g_-(f_0)$. In this case, the JEM $m\rightarrow 0$. Therefore, the dynamics becomes completely a sinusoidal function with finite frequency
\begin{align}
	&\lim_{g_1\rightarrow g_+}\Omega=(f_0^2-1)^{\frac{1}{4}}\sqrt{2(g_+(f_0)-g_2)},\label{eqn:g1g+}
	\\
	&\lim_{g_2\rightarrow g_-}\Omega=(f_0^2-1)^{\frac{1}{4}}\sqrt{2(g_1-g_-(f_0))},\label{eqn:g2g-}
\end{align}
as depicted in Fig.~\ref{fig:example121}(a-b). 
It is worthy noting that as $g_1\rightarrow g_+(f_0)$ or  equivalently $f_0 \rightarrow (g_1+1/g_1)/2$, the classical spin is slightly shaking around the maximum of $f_0$ as time goes on, see Fig.~(\ref{fig:example121})(a). 

As in \textbf{Reg.~I}~(\ref{sec:solu-I}), the JEM $m\rightarrow1$ when $f_0\rightarrow 1$, similar exponential growth of classical dynamics occurs, see Fig.~(\ref{fig:example11})(d) and Fig.~(\ref{fig:example121})(c). However, distinct feature still needs to be emphasized around the saddle point $f_0=1$ ($\theta_0=\pi/2$ and $\phi_0=0$, see Appendix~\ref{Appendix:C}). Although $\Omega\rightarrow 0$ when $f_0\rightarrow1$, we find that the frequency $\Omega$ suffers from a discontinuity as $f_0$ crossing $f_0=1$. For instance, $\Omega(4,-2,1-10^{-6})\approx0.521$ and $\Omega(4,-2,1+10^{-6})\approx1.042$. This discontinuity becomes more clear if we calculate the limit,
\begin{eqnarray}
	\lim_{\epsilon\rightarrow0^+}\frac{ \Omega(g_1,g_2,1-\epsilon)}{\Omega(g_1,g_2,1+\epsilon)}=\frac{1}{2},
	\label{frequency-jump}
\end{eqnarray}
which corresponds to the sharp change of trajectory topology because two fixed points emerge. This phenomenon is also termed as DPT already observed in experiment~\cite{muniz2020,xu2020}. 
At last, 
it would be easily proven that whatever value of $f_0$ ($f_0>1$) is, the JEM $m\rightarrow 1$ when both nonlinear interaction $g_1$ and $g_2$ are strong enough  ($g_1\rightarrow +\infty$ and $g_2\rightarrow -\infty$). As expected, an exponential divergence of classical dynamics occurs with vanishing $\Omega$ whatever the initial state is (Fig.~(\ref{fig:example121})(d)).

\subsubsection{\textbf{Reg.~III}: $g_+(f_0)\leq g_2\leq g_1$}\label{sec:solu-III}
As shown in Table.~\ref{tab:table1}, $\Delta\geq 0$ and $u\geq 0$ and $\sqrt{u^2/4-v}<|u|/2=u/2$ in this regime.
By introducing $\tau^\prime = i\xi w_+\tau$ (Eq.~(\ref{eqn:tau}), and $\xi=\pm 1$ determined by the initial conditions), $w_\pm = \sqrt{u/2\pm \sqrt{\Delta}}$, $Y=X/w_-$ and the JEM $m=w_-^2/w_+^2$, 
\begin{eqnarray}
	m=\frac{D-\sqrt{R_1R_2}}{D+\sqrt{R_1R_2}},
\end{eqnarray}
we discover that Eq.~(\ref{eqn:X}) can be rewritten as follow, 
\begin{equation}
	(\partial_{\tau^\prime} Y)^2=(1+Y^2)(1+m Y^2), 
	\label{eqn:Y3}
\end{equation}
which is identical to the differential equation of sc$(z,m)$~\cite{armitage2006}.
Therefore the solution of Eq.~(\ref{eqn:X})  
\begin{eqnarray}
	X(t)=iw_-\cdot {\rm sn}(\xi w_+\tau+p_0,m),
	\label{eqn:solution122}
\end{eqnarray}
here the imaginary transformation of JEFs, $i\cdot{\rm sn}(u,m)={\rm sc}(iu,1-m)$ has been applied. Similar to the previous cases, $p_0$ and $\xi$ are determined by the initial conditions, 
\begin{eqnarray}
	X(0)&=&iw_-\cdot {\rm sn}(p_0,m),\\
	X^\prime(0) &\propto& i\xi {\rm cn}(p_0,m)\cdot{\rm dn}(p_0,m).
\end{eqnarray} 
The temporal evolution of $X(t)$ can be viewed as a trajectory that parallels to the real axis in complex plane because $p_0\in \mathcal{C}$ with the frequency
\begin{equation}
	\Omega(g_1,g_2,f_0)=\frac{\pi}{2\sqrt{2}K}\sqrt{\sqrt{R_1R_2}+D}.
\end{equation}

\begin{figure}
	\includegraphics[width=0.95\linewidth]{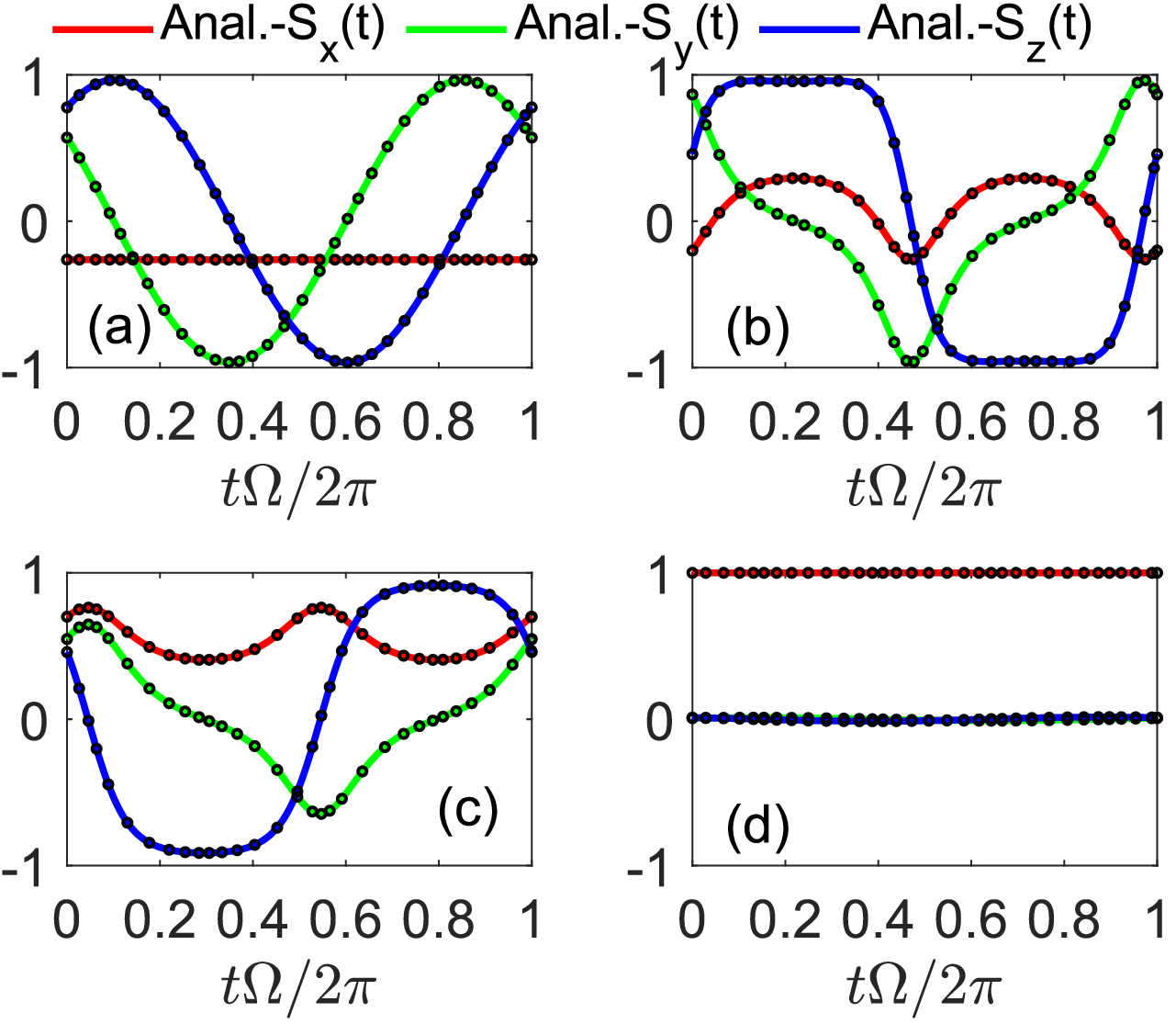}
	\caption{(color online) 
		Comparisons between the analytical solutions Eq.~(\ref{eqn:solution122}) and the numerical ones Eq.~(\ref{eqn:Sxyz}). The solid red, green and blue line corresponds to the $S_x(t)$, $S_y(t)$ and $S_z(t)$, respectively. While the black circles are the corresponding results obtained by numerically solving Eq.~(\ref{eqn:Sxyz}). For subplot (a-d), 
		(a) $g_1=4$, $g_2=4-\epsilon$, and $f_0=1.6$; 
		(b-c), $g_1=4$ $g_2=g_+(f_0)+\epsilon$, and  $f_0=1.6$ (with different initial states); 
		(d), $g_1=20$, $g_2=11$, and $f_0=1+\epsilon$. Here $\epsilon$ is an infinitesimal positive real number.}
	\label{fig:example122}
\end{figure}

As $g_1$ approaches to $g_2$ ($g_1\rightarrow g$ and $g_2\rightarrow g$), the JEM $m\rightarrow 0$, the dynamics behave the same as in \textbf{Reg.~I}~(Sec.~\ref{sec:solu-I}) with frequency $\Omega\rightarrow\sqrt{g^2-2f_0g+1}$ (Fig.~\ref{fig:example11}(b) and Fig.~\ref{fig:example122}(a)). 
In contrast to the behavior around the saddle-point $f_0=1$ in \textbf{Reg.~II}~(Sec.~\ref{sec:solu-II}), however, we find the frequency $\Omega$ is continuous around the saddle point $f_0=(g_2+1/g_2)/2$,
\begin{equation}
	\lim_{\epsilon\rightarrow 0^+}\Omega(g_1,g_++\epsilon,f_0)=\lim_{\epsilon \rightarrow 0^+}\Omega(g_1,g_+-\epsilon,f_0)=0.
	\label{frequency-continuous-1}
\end{equation}
This continuity means that as $g_2$ crossing $g_+(f_0)$,
three different trajectories, i.e., see Fig.~\ref{fig:example122}(b,c) and Fig.~\ref{fig:example212}(b) (that will be discussed in next subsection), approximately share the same frequency because all of them envelop even number of saddle-points. 
Further investigation is needed to uncover the corresponding properties of these three distinct trajectories in relation to coherent atomic oscillations~\cite{raghavan1999}.

Furthermore, because $f_0=1$ in current \textbf{Reg.~III}~(Sec.~\ref{sec:solu-III}) is a local minimum (Fig.~(\ref{fig:landscape-energy})(g)),  
a similar frequency continuity still holds around $f_0=1$, 
\begin{align}
	\lim_{f_0 \rightarrow1^+}\Omega(g_1,g_2,f_0)&=\lim_{f_0\rightarrow1^-}\Omega(g_1,g_2,f_0)\nonumber\\
	&=\sqrt{(g_1-1)(g_2-1)},
	\label{frequency-continuous-2}
\end{align} 
which corresponds to distinct trajectories on the surface of Bloch sphere, respectively. In relation to a BEC realization, the former trajectory around Eq.~(\ref{frequency-continuous-2}) is the plasma oscillation around $(S_x,S_y,S_z)=(1,0,0)$, while the later is the $\pi$-phase oscillation around $(S_x,S_y,S_z)=(-1,0,0)$, please see Fig.~\ref{fig:example11}(c) and Fig.~\ref{fig:example122}(d).

\subsubsection{\textbf{Reg.~IV}: $g_-(f_0)\leq g_2\leq g_+(f_0)$}\label{sec:solu-IV}
By introducing $t_-=-u/2+\sqrt{v}$, $w_\pm =\left(1\pm\sqrt{t_-/(2\sqrt{v})}\right)/2$, $Y=\sqrt{w_+}(X-v^{\frac{1}{4}})/(\sqrt{w_-}(X+v^{\frac{1}{4}}))$, $\tau^\prime = 2i\xi v^{1/4}w_+\tau$~(Eq.~(\ref{eqn:tau})), and the JEM $m=(w_-/w_+)^2$, 
\begin{eqnarray}
	m=\left(\frac{1-\sqrt{\frac{1}{2}\left(1-\frac{D}{(g_1-g_2)\sqrt{f_0^2-1}}\right)}}{1+\sqrt{\frac{1}{2}\left(1-\frac{D}{(g_1-g_2)\sqrt{f_0^2-1}}\right)}}\right)^2, 
\end{eqnarray}
we find that the differential equation of $Y$ can be recast as follow,  
\begin{eqnarray}
	(\partial_{\tau^\prime} Y)^2=(Y^2+1)((1-m)Y^2+1),
	\label{eqn:Y4}
\end{eqnarray}
which is identical to the differential equation of JEF sz$(z,m)$. Hence, the analytical formula of $Y(t)$ can be expressed as follow,
\begin{eqnarray}
	Y(t)=i\sqrt{\frac{w_-}{w_+}}\cdot {\rm sn}(2\xi v^{\frac{1}{4}}b_+\tau+p_0,m),
\end{eqnarray}
where $p_0\in\mathcal{C}$ and $\xi=\pm1$ are determined by the initial conditions, 
\begin{eqnarray}
	Y(0)=\frac{X(0)-\alpha}{X(0)+\alpha}=i\sqrt{\frac{w_-}{w_+}}\cdot {\rm sn}(p_0,m),
\end{eqnarray}
and
\begin{eqnarray}
	\frac{\partial Y(t)}{\partial t}\Big|_{t=0}
	\propto i\xi\cdot {\rm cn}(p_0,m)\cdot{\rm dn}(p_0,m).
\end{eqnarray}
Finally, the analytical solution of $X(t)$ is obtained
\begin{eqnarray}
	X(t)=v^{\frac{1}{4}}\frac{1+Y(t)}{1-Y(t)},
	\label{eqn:solution212}
\end{eqnarray}
with frequency,
\begin{align}
	\Omega
	=\frac{\pi}{4K}&\Big(\sqrt{2}\sqrt{\sqrt{f_0^2-1}\cdot|g_1-g_2|}\nonumber\\
	&+\sqrt{\sqrt{f_0^2-1}\cdot|g_1-g_2|-D}\Big).
\end{align}

\begin{figure}
	\includegraphics[width=\linewidth]{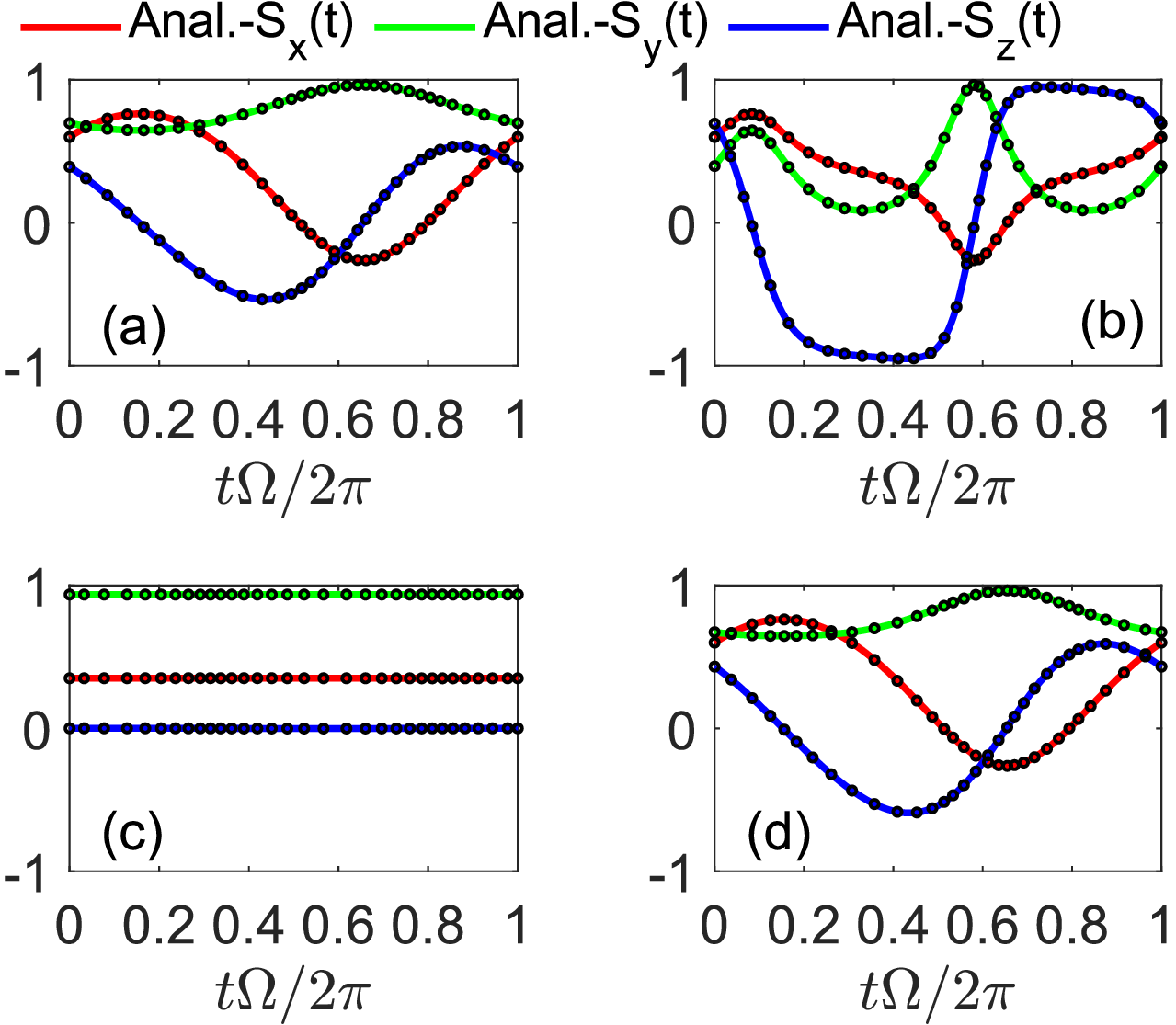}
	\caption{(color online) 
		Comparisons between the analytical solutions Eq.~(\ref{eqn:solution212}) and the numerical ones Eq.~(\ref{eqn:Sxyz}). The solid red, green and blue line corresponds to the $S_x(t)$, $S_y(t)$ and $S_z(t)$, respectively. While the black circles are the corresponding results obtained by numerically solving Eq.~\ref{eqn:Sxyz}.  
		For subplot (a-d), 
		(a) $g_1=4$ and $g_2=g_-(f_0)+\epsilon$; 
		(b), $g_1=4$ and $g_2=g_+(f_0)-\epsilon$; 
		(c), $g_1=g_+(f_0)+\epsilon$ and $g_2=1$; 
		(d), $g_1=4$ and $g_2=1$. Here $f_0=1.6$ and $\epsilon$ is an infinitesimal positive  real number.}
	\label{fig:example212}
\end{figure}

Because frequency is continuous around the boundary $g_2= g_\pm(f_0)$ as already discussed in \textbf{Reg.~II} ~(Sec.~\ref{sec:solu-II}) and \textbf{Reg.~III} ~(Sec.~\ref{sec:solu-III}). Here we just present the corresponding temporal evolution of $S_{x,y,z}(t)$, see Fig.~(\ref{fig:example212})(a,b). A note is that Fig.~(\ref{fig:example212})(b) corresponds to the third trajectory sharing the same frequency as discussed in Eq.~(\ref{frequency-continuous-1}). 
In case $f_0>1$, 
the corresponding frequency is the same formula as in Eq.~(\ref{eqn:g1g+}) when $g_1\rightarrow g_+(f_0)$.  And a similar sinusoidal oscillation around $f_0=(g_1+1/g_1)/2$ is present in Fig.~(\ref{fig:example212})(c). 

\section{}\label{Appendix:A}
The constructions of $S_{x,y,z}(t)$ based on the solutions $X(t)$ (Eq.~(\ref{eqn:X})) hinges on the signs of $g_1$ and $g_2$, therefore, four cases are listed below. 

Case I: when $g_1>0$ and $g_2<0$, set $X_\pm=a_1S_y\pm a_2S_z$, the solutions can be written down as
\begin{eqnarray}
	S_y= \frac{X_+ + X_-}{2a_1},\ 
	S_z= \frac{X_+ - X_-}{2a_2}.
\end{eqnarray}
Case II: when $g_1<0$ and $g_2>0$, set $X_\pm=\pm a_1S_y+ a_2S_z$, 
\begin{eqnarray}
	S_y= \frac{X_+ - X_-}{2a_1},\ 
	S_z= \frac{X_+ + X_-}{2a_2}.
\end{eqnarray}
Case III: when $g_1>g_2>0$ or $g_1<g_2<0$,  
\begin{eqnarray}
	S_y= \frac{X+X^*}{2a_1},\ S_z= \frac{X-X^*}{2a_2}. 
\end{eqnarray}
Case IV: $0<g_1<g_2$ or $g_2<g_1<0$,
\begin{eqnarray}
	S_y= \frac{X-X^*}{2a_1},\ S_z= \frac{X+X^*}{2a_2}.
\end{eqnarray}
Finally, $S_x(t)$ is obtained accordingly, 
\begin{equation}
	S_x=f_0-\frac{1}{2}\big( g_1 S_y^2+g_2 S_z^2 \big)
\end{equation}

\section{}\label{Appendix:B}
In this Appendix, we will present how to solve a equation~\cite{wangzhuxi1989} like, 
\begin{eqnarray}
	(d_t X)^2=P[X]=\sum_{j=0}^4C_jX^j,
	\label{eqn:appendix:dtPX}
\end{eqnarray}
with real coefficients $C_j$. 
%with $\xi^\prime=\pm 1$ which is determined by the initial conditions, $\frac{dX}{dt}|_{t=0}=\xi^\prime \sqrt{P[X]}$. 

Supposing that the four roots of $P[X]=0$, i.e., $a$, $b$, $c$ and $d$, are already known, $P[X]$ can be factorized as follow,
\begin{eqnarray}
	P[X]=S_1S_2,
\end{eqnarray}
with 
\begin{eqnarray}
	&S_1=p_1(X-a)(X-b)=p_1X^2+2q_1X+r_1,\nonumber\\
	&S_2=p_2(X-c)(X-d)=p_2X^2+2q_2X+r_2.\nonumber\\
	\label{eqn:appendix:S1S2}
\end{eqnarray}
Therefore, $p_1=1$, and $p_2=C_4$,  
\begin{eqnarray}
	q_1=-\frac{p_1}{2}(a+b),\ r_1=p_1ab,\nonumber\\
	q_2=-\frac{p_2}{2}(c+d),\ r_2=p_2cd.
	\label{eqn:appendix:coeff}
\end{eqnarray}
Next, we look for a parameter $\lambda$ such that
\begin{eqnarray}
	S_1-\lambda S_2&=&(p_1-\lambda p_2)X^2+2(q_1-\lambda q_2)X\nonumber\\	&&+(r_1-\lambda r_2)\nonumber\\
	&=&(p_1-\lambda p_2)(X-\alpha)^2,
	\label{eqn:appendix:S1S2-2}
\end{eqnarray}
therefore, the discriminant of above quadratic polynomial with variable $X$ must be zero, 
\begin{eqnarray}
	(q_1-\lambda q_2)^2-(p_1-\lambda p_2)(r_1-\lambda r_2)=0.
	\label{eqn:appendix:lambda}
\end{eqnarray}
In general, Eq.~(\ref{eqn:appendix:lambda}) would have two different roots, the another is labeled as $\mu$. While $\alpha$ (the root of $S_1-\lambda S_2$) can be obtained by comparing the coefficient of $X$ in both sides of Eq.(\ref{eqn:appendix:S1S2-2}), as a result,
\begin{eqnarray}
	\lambda = \frac{\alpha p_1 +q_1}{\alpha p_2+q_2}.
	\label{eqn:appendix:lambda-2}
\end{eqnarray}
Substituting $\lambda$ of Eq.~(\ref{eqn:appendix:lambda}) with Eq.~(\ref{eqn:appendix:lambda-2}), we obtain
\begin{align}
	(p_1q_2-p_2q_1)&\cdot\big((p_1q_2-p_2q_1)\alpha^2\nonumber\\
	&+(p_1r_2-p_2r_1)\alpha+q_1r_2-q_2r_1\big)=0,
	\label{eqn:appendix:criteria}
\end{align}
with 
\begin{eqnarray}
	p_1q_2-p_2q_1&=&-\frac{p_1p_2}{2}(c+d-a-b),\nonumber\\
	p_1r_2-p_2r_1&=&p_1p_2(cd-ab),\nonumber\\
	q_1r_2-q_2r_1&=&-\frac{p_1p_2}{2}((a+b)cd-(c+d)ab).\nonumber\\
\end{eqnarray}

Two cases have to be discussed. The first case is  $p_1q_2=p_2q_1$, which results from $a+b=c+d$. $P[X]$ is written down as follow,
\begin{align}
	P[X]&=p_1p_2\cdot\nonumber\\
	&((X-\alpha)^2+ab-\alpha^2)((X-\alpha)^2+cd-\alpha^2),
\end{align}
with $\alpha=(a+b)/2$.
Therefore, Eq.~(\ref{eqn:appendix:dtPX}) becomes
\begin{eqnarray}
	(d_t X)^2 =p_1p_2((X-\alpha)^2+ab-\alpha^2)\nonumber\\
	\cdot((X-\alpha)^2+cd-\alpha^2).\nonumber\\
	\label{eqn:appendix:case1}
\end{eqnarray}

While the second case $p_1q_2\neq p_2q_1$ would be a little complex. The discriminant of Eq.~(\ref{eqn:appendix:criteria}) $\tilde\Delta$ can be simplified as follow, 
\begin{eqnarray}
	\tilde\Delta&=&(p_1r_2-p_2r_1)^2-4(p_1q_2-p_2q_1)(q_1r_2-q_2r_1),\nonumber\\
	&=&(p_1p_2)^2(a-c)(b-c)(a-d)(b-d),
\end{eqnarray}
which is a real and positive number, namely $\tilde\Delta>0$, proven by dividing it into three cases:
Case (1),  $\{a,b,c,d\}\in\mathcal{R}$, and without loss of generality, we can choose $a>b>c>d$, therefore, $\tilde\Delta>0$. 
Case (2), one pair root of $P[X]=0$ are complex numbers while the another are real ones, i.e., $c=d^*$ while $\{a,b\}\in\mathcal{R}$, then $\tilde\Delta=(p_1p_2)^2|a-c|^2\cdot|b-c^*|^2>0$. 
Case (3), all roots of $P[X]=0$ are complexes,  then $\tilde\Delta=(p_1p_2)^2|a-c|^2\cdot |a^*-c|^2>0$. 

Supposing that the two real roots of Eq.~(\ref{eqn:appendix:criteria}) are labeled as $\alpha$ and $\beta$ ($\alpha>\beta$ is assumed without loss of generality). 
When $p_1q_2>p_2q_1$, 
\begin{eqnarray}
	\alpha=\frac{p_1p_2(ab-cd)+\sqrt{\tilde\Delta}}{p_1p_2(a+b-c-d)},\\
	\beta=\frac{p_1p_2(ab-cd)-\sqrt{\tilde\Delta}}{p_1p_2(a+b-c-d)},
\end{eqnarray}
otherwise
\begin{eqnarray}
	\alpha=\frac{p_1p_2(ab-cd)-\sqrt{\tilde\Delta}}{p_1p_2(a+b-c-d)},\\
	\beta=\frac{p_1p_2(ab-cd)+\sqrt{\tilde\Delta}}{p_1p_2(a+b-c-d)}.
\end{eqnarray}
and consequently 
\begin{eqnarray}
	\lambda=\frac{p_1\alpha+q_1}{p_2\alpha+q_2},\nonumber\\ \mu=\frac{p_1\beta+q_1}{p_2\beta+q_2}.
\end{eqnarray}
Substituting the above equations into Eq.~(\ref{eqn:appendix:S1S2}), we finally get, 
\begin{eqnarray}
	&S_1=b_1(X-\alpha)^2+c_1(X-\beta)^2,\nonumber\\
	&S_2=b_2(X-\alpha)^2+c_2(X-\beta)^2,
\end{eqnarray}
with
\begin{eqnarray}
	\begin{array}{cc}
		b_1=\frac{p_1\beta+q_1}{\beta-\alpha}, & c_1=\frac{p_1\alpha+q_1}{\alpha-\beta},\\
		b_2=\frac{p_2\beta+q_2}{\beta-\alpha}, & c_2=\frac{p_2\alpha+q_2}{\alpha-\beta}.
	\end{array}
\end{eqnarray}
Making a transformation, $Z=\frac{X-\alpha}{X-\beta}$, we get
\begin{equation}
	P[X]=\frac{(\alpha-\beta)^4}{(1-Z)^4}(b_1Z^2+c_1)(b_2Z^2+c_2),
\end{equation}
therefore,
\begin{eqnarray}
	(\partial_t Z)^2=(\alpha-\beta)^2(b_1Z^2+c_1)(b_2Z^2+c_2).\nonumber\\
	\label{eqn:appendix:case2}
\end{eqnarray}
Up to a proportional constant, 
the differential equation of $X$ (Eq.~\ref{eqn:appendix:case1}) or $Z$ (Eq.~\ref{eqn:appendix:case2}) behaves identically with one member of family of Jacobi elliptic functions. Finally, the exact behavior of $X$ would be obtained, 
\begin{eqnarray}
	X=\frac{\beta Z-\alpha}{Z-1}.
\end{eqnarray}

\section{}\label{Appendix:C}
\begin{figure}
	\includegraphics[width=0.95\linewidth]{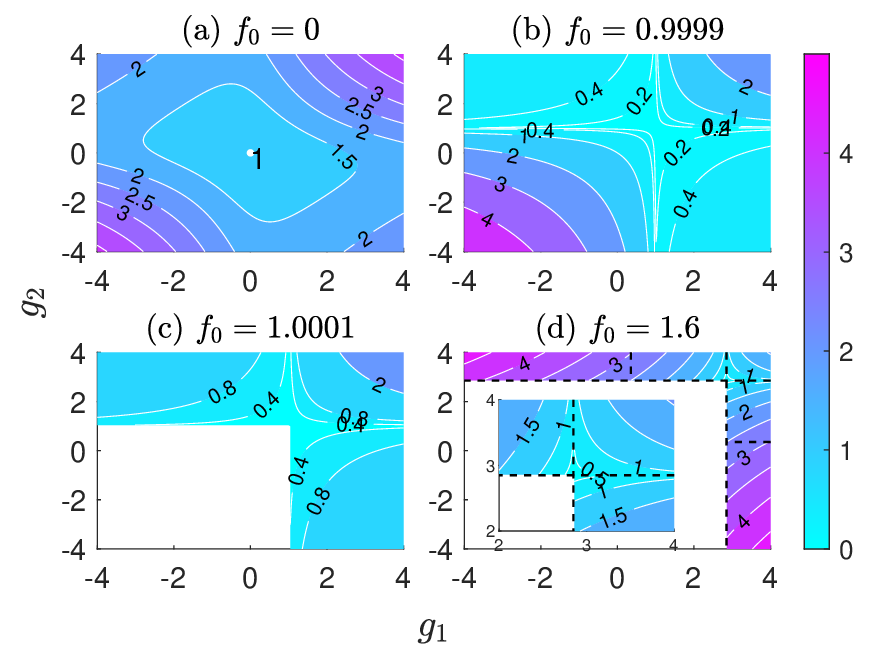}
	\caption{(color online) Frequency $\Omega$ versus $g_1$ and $g_2$ as varying $f_0=0$(a), $0.9999$(b), $1.0001$(c) and $1.6$(d). A zoom-in around $g_1=g_2=g_+(f_0)$ is shown in inset in Fig.~\ref{fig:frequency_f0}(d).}
	\label{fig:frequency_f0}
\end{figure}

In this appendix, we firstly demonstrate the oscillation frequency $\Omega$ for arbitrary nonlinear couplings at a fixed $f_0$ in thermodynamic limit (Fig.~\ref{fig:frequency_f0}), which is helpful to uncover DPT of generic LMG model (Eq.~(\ref{eq:Hamiltonian0})). 
Because $f_0=1$ is a critical point as figured out in the main text, two cases are divided. The first case is $0\leq f_0\leq 1$. Basically, the oscillation frequency $\Omega$ increases as far away from the dip zone around  $(g_1,g_2)=(0,0)$ when  $f_0\approx 0$. And the dip zone moves gradually to the regime around $(g_1,g_2)=(1,1)$ as $f_0\rightarrow 1$, as depicted in Fig.~(\ref{fig:frequency_f0})(a,b), because $f_0=1$ (($S_x$,$S_y$,$S_z$)=(1,0,0)) is a saddle point when $(g_1-1)(g_2-1)<0$. 
The second case is $f_0\geq 1$.  
Eq.~(\ref{eqn:Sxyz}) only exists when $g_1\geq g_+(f_0)$ and $g_1>g_2$, or $g_2\geq g_+(f_0)$ and $g_2>g_1$,  demonstrated in Fig.~(\ref{fig:frequency_f0}))(c,d). Around $f_0=1$, although $\lim_{f_0\rightarrow1}\Omega=0$, a frequency discontinuity can be observed in Fig.~(\ref{fig:frequency_f0})(b,c) (see Eq.~(\ref{frequency-jump})) when $(g_1-1)(g_2-1)<0$. Additionally, $\Omega$ gets additional local minimum around $g_2=g_+(f_0)$, which is proved to be exactly zero (Eq.~(\ref{frequency-continuous-1})). This phenomena originate from a fact that the classical trajectories approach to another saddle point, $f_0=(g_2+1/g_2)/2$. As a comparison, $\Omega$ is finite and continuous around $g_1= g_+(f_0)$ or $g_2= g_-(f_0)$. 

In fact, the spin vector are parameterized by $\theta$ and $\phi$, $S_z(0)=\cos(\theta)$, $S_x(0)=\sin(\theta)\cos(\phi)$, $S_y(0)=\sin(\theta)\sin(\phi)$ with $\theta\in [0,\pi]$ and $\phi\in [0,2\pi]$, then, 
\begin{eqnarray}
	f_0=\sin\theta\cos\phi+\frac{g_1}{2}\sin^2\theta\sin^2\phi+\frac{g_2}{2}\cos^2\theta.\nonumber\\
\end{eqnarray}
The classical dynamics Eq.~{(\ref{eqn:Sxyz})} is greatly dominated by landscape of $f_0$, i.e., the minimum, saddle-point and maximum~\cite{castanos2006}, which are  determined by $\partial_{\theta} f_0=0$ and $\partial_{\phi} f_0=0$,
\begin{eqnarray}
	\cos\theta\big(\cos\phi+\sin\theta(g_1\sin^2\phi-g_2)\big)&=&0,\nonumber\\
	\sin\theta\sin\phi\big(-1+g_1\sin\theta\cos\phi\big)&=&0.\nonumber\\
	\label{eqn:f0-1st}
\end{eqnarray}
In general, for any real values of $g_1$ and $g_2$, six trivial solutions of Eq.~(\ref{eqn:f0-1st}) always exist, 
the first and second solution is $(\theta=\pi/2,\phi=0)$ and  $(\theta=\pi/2,\phi=\pi)$, corresponding to $f_0=1$ and $f_0=-1$ respectively. While the rest four solutions ($\theta=0$ or $\pi$, $\phi=\pi/2$ or $3\pi/2$) share the same energy $f_0=g_2/2$. We note that the classical energy surface obtained here coincides with the one based on spin-coherent state~\cite{Ribeiro2007,Ribeiro2008}. 

\begin{figure}
	\includegraphics[width=0.95\linewidth]{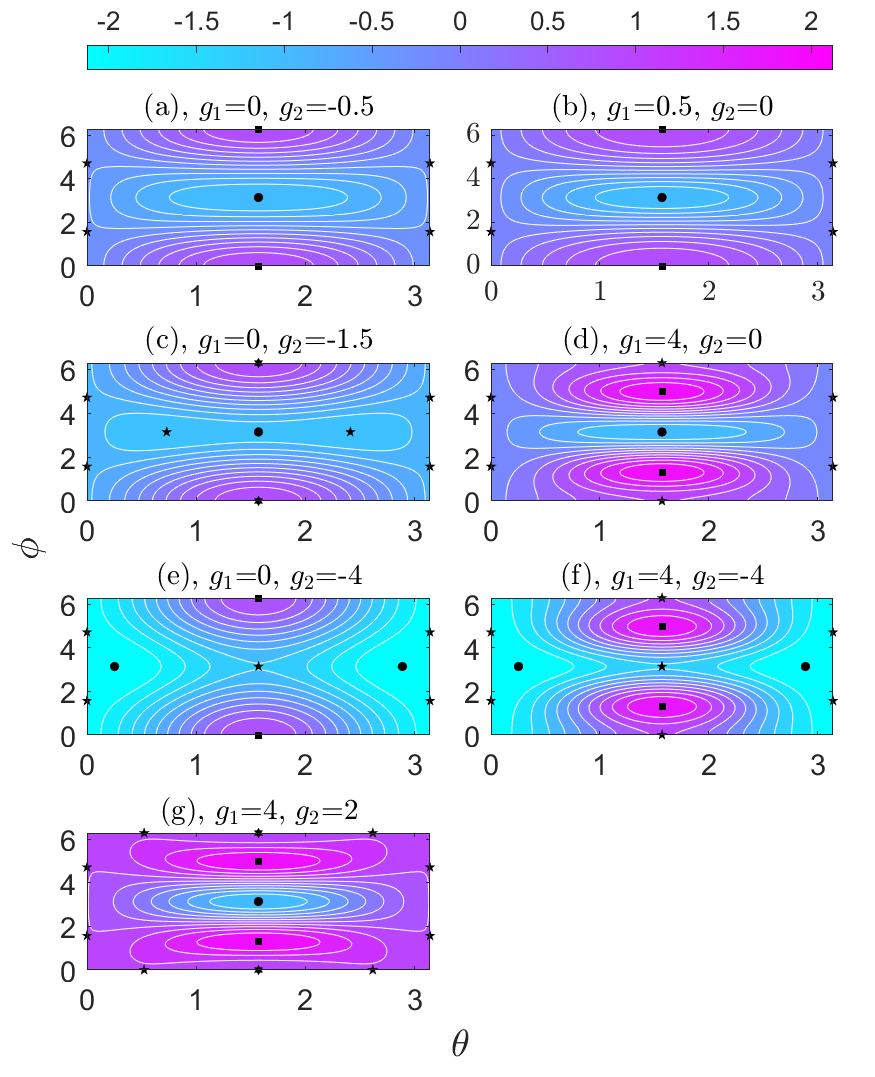}
	\caption{(color online) Landscape of energy, Seven typical examples are chosen to demonstrate the frequency landscape as varying $\theta$ and $\phi$. The global/local minimums, saddle points and global/local maximums are labeled by black circle/hexagon, pentagram and square/diamond symbols, respectively.}\label{fig:landscape-energy}
\end{figure}

Besides these trivial solutions, additional five types of solutions are needed further to be discussed as follow,

The first solution is $\theta=\pi/2$, $\phi=0$, then $f_0=1$, the Hessian matrix, 
\begin{eqnarray}
	M=\left(
	\begin{array}{cc}
		g_2-1 & 0 \\
		0 & g_1-1\\
	\end{array}
	\right).
\end{eqnarray}
Therefore, this solution is a local minimum (maximum) when $g_1\geq1$ and $g_2\geq1$ ($g_1\leq1$ and $g_2\leq1$), which shows in Fig.~(\ref{fig:landscape-energy})(a, b, c, e, g). 
While in the rest regime ($(g_1-1)(g_2-1)<0$), it becomes a saddle point (Fig.~(\ref{fig:landscape-energy})(d,f)).  
The second solution $\theta=\pi/2$ and $\phi=\pi$, $f_0=-1$ and the Hessian matrix,
\begin{eqnarray}
	M=\left(
	\begin{array}{cc}
		g_2+1 & 0 \\
		0 & g_1+1\\
	\end{array}
	\right),
\end{eqnarray}
a similar conclusion is given besides that the critical line becomes  $g_1=-1$ and $g_2=-1$. 

The third solution includes four sub-solutions, $\theta=0$ (or $\pi$) and $\phi=\pi/2$ (or $3\pi/2$), sharing the same energy $f_0=g_2/2$ and eigenvalues of its corresponding Hessian matrix, $E_\pm=\frac{g_1-g_2}{2}\pm\sqrt{\left(\frac{g_1-g_2}{2}\right)^2+1}$. Obviously, $E_-<0$ and $E_+>0$ always hold for any value of $g_1$ and $g_2$. Therefore, the four solution always are saddle points, see the pentagram symbols in Fig.~(\ref{fig:landscape-energy}).

However, two additional solutions emerge when $|g_1|\geq 1$ or $|g_2|\geq 1$. 
The fourth is $\theta=\pi/2$, $\phi=\arccos(1/g_1)$, $f_0=(g_1+1/g_1)/2$, and the Hessian matrix, 
\begin{eqnarray}
	M=\left(
	\begin{array}{cc}
		g_2-g_1 & 0 \\
		0 & \frac{1}{g_1}-g_1\\
	\end{array}
	\right).
\end{eqnarray}
While the fifth is $\phi=0$, $\theta=\arcsin(1/g_2)$ when $g_2>1$, and $\phi=\pi$, $\theta=\arcsin(-1/g_2)$ when $g_2<-1$, the Hessian matrix,
\begin{eqnarray}
	M=\left(
	\begin{array}{cc}
		\frac{1}{g_2}-g_2 & 0 \\
		0 & \frac{g_1-g_2}{g_2^2}\\
	\end{array}
	\right).
\end{eqnarray}
Due to the time-reversal-like symmetry of Eq.~(\ref{eqn:Sxyz}), half $g_1-g_2$ 
plane is only needed to be discussed, i.e., $g_1\geq g_2$. Therefore, the properties of last two solution can be summarized as follow, the fourth one
$\theta=\pi/2$ and $\phi=\arccos(1/g_1)$ is a local maximum (saddle point) when $g_1>1$ ($g_1<-1$), which converges into the first solution at $g_1=\pm1$. While the fifth one $\phi=0$ and $\theta=\arcsin(1/g_2)$ when $g_2\geq 1$ ($\phi=\pi$ and $\theta=\arcsin(-1/g_2)$ when $g_2\leq -1$) is a saddle point (local minimum), which converges into the second solution around $g_2=\pm 1$, see Fig.~(\ref{fig:landscape-energy}).
	
\end{appendices}

\bibliography{sn-bibliography}% common bib file


\begin{thebibliography}{46}
\ifx \bisbn   \undefined \def \bisbn  #1{ISBN #1}\fi
\ifx \binits  \undefined \def \binits#1{#1}\fi
\ifx \bauthor  \undefined \def \bauthor#1{#1}\fi
\ifx \batitle  \undefined \def \batitle#1{#1}\fi
\ifx \bjtitle  \undefined \def \bjtitle#1{#1}\fi
\ifx \bvolume  \undefined \def \bvolume#1{\textbf{#1}}\fi
\ifx \byear  \undefined \def \byear#1{#1}\fi
\ifx \bissue  \undefined \def \bissue#1{#1}\fi
\ifx \bfpage  \undefined \def \bfpage#1{#1}\fi
\ifx \blpage  \undefined \def \blpage #1{#1}\fi
\ifx \burl  \undefined \def \burl#1{\textsf{#1}}\fi
\ifx \doiurl  \undefined \def \doiurl#1{\url{https://doi.org/#1}}\fi
\ifx \betal  \undefined \def \betal{\textit{et al.}}\fi
\ifx \binstitute  \undefined \def \binstitute#1{#1}\fi
\ifx \binstitutionaled  \undefined \def \binstitutionaled#1{#1}\fi
\ifx \bctitle  \undefined \def \bctitle#1{#1}\fi
\ifx \beditor  \undefined \def \beditor#1{#1}\fi
\ifx \bpublisher  \undefined \def \bpublisher#1{#1}\fi
\ifx \bbtitle  \undefined \def \bbtitle#1{#1}\fi
\ifx \bedition  \undefined \def \bedition#1{#1}\fi
\ifx \bseriesno  \undefined \def \bseriesno#1{#1}\fi
\ifx \blocation  \undefined \def \blocation#1{#1}\fi
\ifx \bsertitle  \undefined \def \bsertitle#1{#1}\fi
\ifx \bsnm \undefined \def \bsnm#1{#1}\fi
\ifx \bsuffix \undefined \def \bsuffix#1{#1}\fi
\ifx \bparticle \undefined \def \bparticle#1{#1}\fi
\ifx \barticle \undefined \def \barticle#1{#1}\fi
\bibcommenthead
\ifx \bconfdate \undefined \def \bconfdate #1{#1}\fi
\ifx \botherref \undefined \def \botherref #1{#1}\fi
\ifx \url \undefined \def \url#1{\textsf{#1}}\fi
\ifx \bchapter \undefined \def \bchapter#1{#1}\fi
\ifx \bbook \undefined \def \bbook#1{#1}\fi
\ifx \bcomment \undefined \def \bcomment#1{#1}\fi
\ifx \oauthor \undefined \def \oauthor#1{#1}\fi
\ifx \citeauthoryear \undefined \def \citeauthoryear#1{#1}\fi
\ifx \endbibitem  \undefined \def \endbibitem {}\fi
\ifx \bconflocation  \undefined \def \bconflocation#1{#1}\fi
\ifx \arxivurl  \undefined \def \arxivurl#1{\textsf{#1}}\fi
\csname PreBibitemsHook\endcsname

%%% 1
\bibitem[\protect\citeauthoryear{Lipkin et~al.}{1965}]{lipkin1965}
\begin{barticle}
\bauthor{\bsnm{Lipkin}, \binits{H.J.}},
\bauthor{\bsnm{Meshkov}, \binits{N.}},
\bauthor{\bsnm{Glick}, \binits{A.J.}}:
\batitle{Validity of many-body approximation methods for a solvable model: (i).
  exact solutions and perturbation theory}.
\bjtitle{Nuclear Physics}
\bvolume{62}(\bissue{2}),
\bfpage{188}--\blpage{198}
(\byear{1965})
\doiurl{10.1016/0029-5582(65)90862-X}
\end{barticle}
\endbibitem

%%% 2
\bibitem[\protect\citeauthoryear{Meshkov et~al.}{1965}]{meshkov1965}
\begin{barticle}
\bauthor{\bsnm{Meshkov}, \binits{N.}},
\bauthor{\bsnm{Glick}, \binits{A.J.}},
\bauthor{\bsnm{Lipkin}, \binits{H.J.}}:
\batitle{Validity of many-body approximation methods for a solvable model:
  (ii). linearization procedures}.
\bjtitle{Nuclear Physics}
\bvolume{62}(\bissue{2}),
\bfpage{199}--\blpage{210}
(\byear{1965})
\doiurl{10.1016/0029-5582(65)90863-1}
\end{barticle}
\endbibitem

%%% 3
\bibitem[\protect\citeauthoryear{Glick et~al.}{1965}]{glick1965}
\begin{barticle}
\bauthor{\bsnm{Glick}, \binits{A.J.}},
\bauthor{\bsnm{Lipkin}, \binits{H.J.}},
\bauthor{\bsnm{Meshkov}, \binits{N.}}:
\batitle{Validity of many-body approximation methods for a solvable model:
  (iii). diagram summations}.
\bjtitle{Nuclear Physics}
\bvolume{62}(\bissue{2}),
\bfpage{211}--\blpage{224}
(\byear{1965})
\doiurl{10.1016/0029-5582(65)90864-3}
\end{barticle}
\endbibitem

%%% 4
\bibitem[\protect\citeauthoryear{Carrasco et~al.}{2016}]{carrasco2016}
\begin{barticle}
\bauthor{\bsnm{Carrasco}, \binits{J.A.}},
\bauthor{\bsnm{Finkel}, \binits{F.}},
\bauthor{\bsnm{Gonzalez-Lopez}, \binits{A.}},
\bauthor{\bsnm{Rodriguez}, \binits{M.A.}},
\bauthor{\bsnm{Tempesta}, \binits{P.}}:
\batitle{Generalized isotropic lipkin–meshkov–glick models: ground state
  entanglement and quantum entropies}.
\bjtitle{Journal of Statistical Mechanics: Theory and Experiment}
\bvolume{2016}(\bissue{3}),
\bfpage{033114}
(\byear{2016})
\doiurl{10.1088/1742-5468/2016/03/033114}
\end{barticle}
\endbibitem

%%% 5
\bibitem[\protect\citeauthoryear{Raghavan et~al.}{1999}]{raghavan1999}
\begin{barticle}
\bauthor{\bsnm{Raghavan}, \binits{S.}},
\bauthor{\bsnm{Smerzi}, \binits{A.}},
\bauthor{\bsnm{Fantoni}, \binits{S.}},
\bauthor{\bsnm{Shenoy}, \binits{S.R.}}:
\batitle{Coherent oscillations between two weakly coupled bose-einstein
  condensates: Josephson effects, $\ensuremath{\pi}$ oscillations, and
  macroscopic quantum self-trapping}.
\bjtitle{Phys. Rev. A}
\bvolume{59},
\bfpage{620}--\blpage{633}
(\byear{1999})
\doiurl{10.1103/PhysRevA.59.620}
\end{barticle}
\endbibitem

%%% 6
\bibitem[\protect\citeauthoryear{Zibold et~al.}{2010}]{zibold2010}
\begin{barticle}
\bauthor{\bsnm{Zibold}, \binits{T.}},
\bauthor{\bsnm{Nicklas}, \binits{E.}},
\bauthor{\bsnm{Gross}, \binits{C.}},
\bauthor{\bsnm{Oberthaler}, \binits{M.K.}}:
\batitle{Classical bifurcation at the transition from rabi to josephson
  dynamics}.
\bjtitle{Phys. Rev. Lett.}
\bvolume{105},
\bfpage{204101}
(\byear{2010})
\doiurl{10.1103/PhysRevLett.105.204101}
\end{barticle}
\endbibitem

%%% 7
\bibitem[\protect\citeauthoryear{Baumann et~al.}{2010}]{baumann2010}
\begin{barticle}
\bauthor{\bsnm{Baumann}, \binits{K.}},
\bauthor{\bsnm{Guerlin}, \binits{C.}},
\bauthor{\bsnm{Brennecke}, \binits{F.}},
\bauthor{\bsnm{Esslinger}, \binits{T.}}:
\batitle{Dicke quantum phase transition with a superfluid gas in an optical
  cavity}.
\bjtitle{nature}
\bvolume{464}(\bissue{7293}),
\bfpage{1301}--\blpage{1306}
(\byear{2010})
\doiurl{10.1038/nature09009}
\end{barticle}
\endbibitem

%%% 8
\bibitem[\protect\citeauthoryear{Muniz et~al.}{2020}]{muniz2020}
\begin{barticle}
\bauthor{\bsnm{Muniz}, \binits{J.A.}},
\bauthor{\bsnm{Barberena}, \binits{D.}},
\bauthor{\bsnm{Lewis-Swan}, \binits{R.J.}},
\bauthor{\bsnm{Young}, \binits{D.J.}},
\bauthor{\bsnm{Cline}, \binits{J.R.K.}},
\bauthor{\bsnm{Rey}, \binits{A.M.}},
\bauthor{\bsnm{Thompson}, \binits{J.K.}}:
\batitle{Exploring dynamical phase transitions with cold atoms in an optical
  cavity}.
\bjtitle{NATURE}
\bvolume{580}(\bissue{7805}),
\bfpage{602}
(\byear{2020})
\doiurl{10.1038/s41586-020-2224-x}
\end{barticle}
\endbibitem

%%% 9
\bibitem[\protect\citeauthoryear{Chu et~al.}{2020}]{chuAJ2020}
\begin{barticle}
\bauthor{\bsnm{Chu}, \binits{A.}},
\bauthor{\bsnm{Will}, \binits{J.}},
\bauthor{\bsnm{Arlt}, \binits{J.}},
\bauthor{\bsnm{Klempt}, \binits{C.}},
\bauthor{\bsnm{Rey}, \binits{A.M.}}:
\batitle{Simulation of $xxz$ spin models using sideband transitions in trapped
  bosonic gases}.
\bjtitle{Phys. Rev. Lett.}
\bvolume{125},
\bfpage{240504}
(\byear{2020})
\doiurl{10.1103/PhysRevLett.125.240504}
\end{barticle}
\endbibitem

%%% 10
\bibitem[\protect\citeauthoryear{Xu et~al.}{2020}]{xu2020}
\begin{barticle}
\bauthor{\bsnm{Xu}, \binits{K.}},
\bauthor{\bsnm{Sun}, \binits{Z.-H.}},
\bauthor{\bsnm{Liu}, \binits{W.}},
\bauthor{\bsnm{Zhang}, \binits{Y.-R.}},
\bauthor{\bsnm{Li}, \binits{H.}},
\bauthor{\bsnm{Dong}, \binits{H.}},
\bauthor{\bsnm{Ren}, \binits{W.}},
\bauthor{\bsnm{Zhang}, \binits{P.}},
\bauthor{\bsnm{Nori}, \binits{F.}},
\bauthor{\bsnm{Zheng}, \binits{D.}},
\bauthor{\bsnm{Fan}, \binits{H.}},
\bauthor{\bsnm{Wang}, \binits{H.}}:
\batitle{Probing dynamical phase transitions with a superconducting quantum
  simulator}.
\bjtitle{Science Advances}
\bvolume{6}(\bissue{25}),
\bfpage{4935}
(\byear{2020})
\doiurl{10.1126/sciadv.aba4935}
\end{barticle}
\endbibitem

%%% 11
\bibitem[\protect\citeauthoryear{Zhou et~al.}{2024}]{Zhou2024}
\begin{botherref}
\oauthor{\bsnm{Zhou}, \binits{Y.}},
\oauthor{\bsnm{Wang}, \binits{J.-W.}},
\oauthor{\bsnm{Cao}, \binits{L.-Z.}},
\oauthor{\bsnm{Wang}, \binits{G.-H.}},
\oauthor{\bsnm{Shi}, \binits{Z.-Y.}},
\oauthor{\bsnm{Lu}, \binits{D.-Y.}},
\oauthor{\bsnm{Huang}, \binits{H.-B.}},
\oauthor{\bsnm{Hu}, \binits{C.-S.}}:
Realization of chiral two-mode lipkin-meshkov-glick models via acoustics.
REPORTS ON PROGRESS IN PHYSICS
\textbf{87}(10)
(2024)
\doiurl{10.1088/1361-6633/ad797d}
\end{botherref}
\endbibitem

%%% 12
\bibitem[\protect\citeauthoryear{Pezze et~al.}{2018}]{pezze2018}
\begin{barticle}
\bauthor{\bsnm{Pezze}, \binits{L.}},
\bauthor{\bsnm{Smerzi}, \binits{A.}},
\bauthor{\bsnm{Oberthaler}, \binits{M.K.}},
\bauthor{\bsnm{Schmied}, \binits{R.}},
\bauthor{\bsnm{Treutlein}, \binits{P.}}:
\batitle{Quantum metrology with nonclassical states of atomic ensembles}.
\bjtitle{Rev. Mod. Phys.}
\bvolume{90},
\bfpage{035005}
(\byear{2018})
\doiurl{10.1103/RevModPhys.90.035005}
\end{barticle}
\endbibitem

%%% 13
\bibitem[\protect\citeauthoryear{Morita et~al.}{2006}]{morita2006}
\begin{barticle}
\bauthor{\bsnm{Morita}, \binits{H.}},
\bauthor{\bsnm{Ohnishi}, \binits{H.}},
\bauthor{\bsnm{Providencia}, \binits{J.}},
\bauthor{\bsnm{Nishiyama}, \binits{S.}}:
\batitle{Exact solutions for the lmg model hamiltonian based on the bethe
  ansatz}.
\bjtitle{Nuclear Physics B}
\bvolume{737}(\bissue{3}),
\bfpage{337}--\blpage{350}
(\byear{2006})
\doiurl{10.1016/j.nuclphysb.2006.01.015}
\end{barticle}
\endbibitem

%%% 14
\bibitem[\protect\citeauthoryear{Ribeiro et~al.}{2008}]{Ribeiro2008}
\begin{barticle}
\bauthor{\bsnm{Ribeiro}, \binits{P.}},
\bauthor{\bsnm{Vidal}, \binits{J.}},
\bauthor{\bsnm{Mosseri}, \binits{R.}}:
\batitle{Exact spectrum of the lipkin-meshkov-glick model in the thermodynamic
  limit and finite-size corrections}.
\bjtitle{Phys. Rev. E}
\bvolume{78},
\bfpage{021106}
(\byear{2008})
\doiurl{10.1103/PhysRevE.78.021106}
\end{barticle}
\endbibitem

%%% 15
\bibitem[\protect\citeauthoryear{Unanyan and Fleischhauer}{2003}]{Unanyan2003}
\begin{barticle}
\bauthor{\bsnm{Unanyan}, \binits{R.G.}},
\bauthor{\bsnm{Fleischhauer}, \binits{M.}}:
\batitle{Decoherence-free generation of many-particle entanglement by adiabatic
  ground-state transitions}.
\bjtitle{Phys. Rev. Lett.}
\bvolume{90},
\bfpage{133601}
(\byear{2003})
\doiurl{10.1103/PhysRevLett.90.133601}
\end{barticle}
\endbibitem

%%% 16
\bibitem[\protect\citeauthoryear{Or\'us et~al.}{2008}]{Orus2008}
\begin{barticle}
\bauthor{\bsnm{Or\'us}, \binits{R.}},
\bauthor{\bsnm{Dusuel}, \binits{S.}},
\bauthor{\bsnm{Vidal}, \binits{J.}}:
\batitle{Equivalence of critical scaling laws for many-body entanglement in the
  lipkin-meshkov-glick model}.
\bjtitle{Phys. Rev. Lett.}
\bvolume{101},
\bfpage{025701}
(\byear{2008})
\doiurl{10.1103/PhysRevLett.101.025701}
\end{barticle}
\endbibitem

%%% 17
\bibitem[\protect\citeauthoryear{Ma et~al.}{2009}]{Ma2009}
\begin{barticle}
\bauthor{\bsnm{Ma}, \binits{J.}},
\bauthor{\bsnm{Wang}, \binits{X.}},
\bauthor{\bsnm{Gu}, \binits{S.-J.}}:
\batitle{Many-body reduced fidelity susceptibility in lipkin-meshkov-glick
  model}.
\bjtitle{Phys. Rev. E}
\bvolume{80},
\bfpage{021124}
(\byear{2009})
\doiurl{10.1103/PhysRevE.80.021124}
\end{barticle}
\endbibitem

%%% 18
\bibitem[\protect\citeauthoryear{Engelhardt et~al.}{2013}]{Engelhardt2013}
\begin{barticle}
\bauthor{\bsnm{Engelhardt}, \binits{G.}},
\bauthor{\bsnm{Bastidas}, \binits{V.M.}},
\bauthor{\bsnm{Emary}, \binits{C.}},
\bauthor{\bsnm{Brandes}, \binits{T.}}:
\batitle{ac-driven quantum phase transition in the lipkin-meshkov-glick model}.
\bjtitle{Phys. Rev. E}
\bvolume{87},
\bfpage{052110}
(\byear{2013})
\doiurl{10.1103/PhysRevE.87.052110}
\end{barticle}
\endbibitem

%%% 19
\bibitem[\protect\citeauthoryear{Engelhardt et~al.}{2015}]{Engelhardt2015}
\begin{barticle}
\bauthor{\bsnm{Engelhardt}, \binits{G.}},
\bauthor{\bsnm{Bastidas}, \binits{V.M.}},
\bauthor{\bsnm{Kopylov}, \binits{W.}},
\bauthor{\bsnm{Brandes}, \binits{T.}}:
\batitle{Excited-state quantum phase transitions and periodic dynamics}.
\bjtitle{Phys. Rev. A}
\bvolume{91},
\bfpage{013631}
(\byear{2015})
\doiurl{10.1103/PhysRevA.91.013631}
\end{barticle}
\endbibitem

%%% 20
\bibitem[\protect\citeauthoryear{Schliemann}{2015}]{Schliemann2015}
\begin{barticle}
\bauthor{\bsnm{Schliemann}, \binits{J.}}:
\batitle{Coherent quantum dynamics: What fluctuations can tell}.
\bjtitle{Phys. Rev. A}
\bvolume{92},
\bfpage{022108}
(\byear{2015})
\doiurl{10.1103/PhysRevA.92.022108}
\end{barticle}
\endbibitem

%%% 21
\bibitem[\protect\citeauthoryear{Lerose et~al.}{2018}]{lerose2018}
\begin{barticle}
\bauthor{\bsnm{Lerose}, \binits{A.}},
\bauthor{\bsnm{Marino}, \binits{J.}},
\bauthor{\bsnm{Zunkovic}, \binits{B.}},
\bauthor{\bsnm{Gambassi}, \binits{A.}},
\bauthor{\bsnm{Silva}, \binits{A.}}:
\batitle{Chaotic dynamical ferromagnetic phase induced by nonequilibrium
  quantum fluctuations}.
\bjtitle{Phys. Rev. Lett.}
\bvolume{120},
\bfpage{130603}
(\byear{2018})
\doiurl{10.1103/PhysRevLett.120.130603}
\end{barticle}
\endbibitem

%%% 22
\bibitem[\protect\citeauthoryear{Bao et~al.}{2020}]{Bao2020}
\begin{barticle}
\bauthor{\bsnm{Bao}, \binits{J.}},
\bauthor{\bsnm{Guo}, \binits{B.}},
\bauthor{\bsnm{Liu}, \binits{Y.-H.}},
\bauthor{\bsnm{Shen}, \binits{L.-H.}},
\bauthor{\bsnm{Sun}, \binits{Z.-Y.}}:
\batitle{Multipartite nonlocality and global quantum discord in the
  antiferromagnetic lipkin–meshkov–glick model}.
\bjtitle{Physica B: Condensed Matter}
\bvolume{593},
\bfpage{412297}
(\byear{2020})
\doiurl{10.1016/j.physb.2020.412297}
\end{barticle}
\endbibitem

%%% 23
\bibitem[\protect\citeauthoryear{Nader et~al.}{2021}]{Nader2021}
\begin{barticle}
\bauthor{\bsnm{Nader}, \binits{D.J.}},
\bauthor{\bsnm{Gonz\'alez-Rodr\'{\i}guez}, \binits{C.A.}},
\bauthor{\bsnm{Lerma-Hern\'andez}, \binits{S.}}:
\batitle{Avoided crossings and dynamical tunneling close to excited-state
  quantum phase transitions}.
\bjtitle{Phys. Rev. E}
\bvolume{104},
\bfpage{064116}
(\byear{2021})
\doiurl{10.1103/PhysRevE.104.064116}
\end{barticle}
\endbibitem

%%% 24
\bibitem[\protect\citeauthoryear{Romero et~al.}{2022}]{Romero2022}
\begin{barticle}
\bauthor{\bsnm{Romero}, \binits{A.M.}},
\bauthor{\bsnm{Engel}, \binits{J.}},
\bauthor{\bsnm{Tang}, \binits{H.L.}},
\bauthor{\bsnm{Economou}, \binits{S.E.}}:
\batitle{Solving nuclear structure problems with the adaptive variational
  quantum algorithm}.
\bjtitle{Phys. Rev. C}
\bvolume{105},
\bfpage{064317}
(\byear{2022})
\doiurl{10.1103/PhysRevC.105.064317}
\end{barticle}
\endbibitem

%%% 25
\bibitem[\protect\citeauthoryear{Viscondi et~al.}{2009}]{viscondi2009}
\begin{barticle}
\bauthor{\bsnm{Viscondi}, \binits{T.F.}},
\bauthor{\bsnm{Furuya}, \binits{K.}},
\bauthor{\bsnm{De~Oliveira}, \binits{M.}}:
\batitle{Coherent state approach to the cross-collisional effects in the
  population dynamics of a two-mode bose--einstein condensate}.
\bjtitle{Annals of Physics}
\bvolume{324}(\bissue{9}),
\bfpage{1837}--\blpage{1846}
(\byear{2009})
\doiurl{10.1016/j.aop.2009.05.008}
\end{barticle}
\endbibitem

%%% 26
\bibitem[\protect\citeauthoryear{Opatrny et~al.}{2015}]{opatrny2015}
\begin{barticle}
\bauthor{\bsnm{Opatrny}, \binits{T.}},
\bauthor{\bsnm{Kolar}, \binits{M.}},
\bauthor{\bsnm{Das}, \binits{K.K.}}:
\batitle{Spin squeezing by tensor twisting and lipkin-meshkov-glick dynamics in
  a toroidal bose-einstein condensate with spatially modulated nonlinearity}.
\bjtitle{Phys. Rev. A}
\bvolume{91},
\bfpage{053612}
(\byear{2015})
\doiurl{10.1103/PhysRevA.91.053612}
\end{barticle}
\endbibitem

%%% 27
\bibitem[\protect\citeauthoryear{Ma et~al.}{2011}]{ma2011}
\begin{barticle}
\bauthor{\bsnm{Ma}, \binits{J.}},
\bauthor{\bsnm{Wang}, \binits{X.}},
\bauthor{\bsnm{Sun}, \binits{C.P.}},
\bauthor{\bsnm{Nori}, \binits{F.}}:
\batitle{Quantum spin squeezing}.
\bjtitle{Physics Reports}
\bvolume{509}(\bissue{2}),
\bfpage{89}--\blpage{165}
(\byear{2011})
\doiurl{10.1016/j.physrep.2011.08.003}
\end{barticle}
\endbibitem

%%% 28
\bibitem[\protect\citeauthoryear{Micheli et~al.}{2003}]{micheli2003}
\begin{barticle}
\bauthor{\bsnm{Micheli}, \binits{A.}},
\bauthor{\bsnm{Jaksch}, \binits{D.}},
\bauthor{\bsnm{Cirac}, \binits{J.I.}},
\bauthor{\bsnm{Zoller}, \binits{P.}}:
\batitle{Many-particle entanglement in two-component bose-einstein
  condensates}.
\bjtitle{Phys. Rev. A}
\bvolume{67},
\bfpage{013607}
(\byear{2003})
\doiurl{10.1103/PhysRevA.67.013607}
\end{barticle}
\endbibitem

%%% 29
\bibitem[\protect\citeauthoryear{Vidal et~al.}{2004}]{vidal2004}
\begin{barticle}
\bauthor{\bsnm{Vidal}, \binits{J.}},
\bauthor{\bsnm{Palacios}, \binits{G.}},
\bauthor{\bsnm{Aslangul}, \binits{C.}}:
\batitle{Entanglement dynamics in the lipkin-meshkov-glick model}.
\bjtitle{Phys. Rev. A}
\bvolume{70},
\bfpage{062304}
(\byear{2004})
\doiurl{10.1103/PhysRevA.70.062304}
\end{barticle}
\endbibitem

%%% 30
\bibitem[\protect\citeauthoryear{Lerose et~al.}{2019}]{lerose2019}
\begin{barticle}
\bauthor{\bsnm{Lerose}, \binits{A.}},
\bauthor{\bsnm{Zunkovic}, \binits{B.}},
\bauthor{\bsnm{Marino}, \binits{J.}},
\bauthor{\bsnm{Gambassi}, \binits{A.}},
\bauthor{\bsnm{Silva}, \binits{A.}}:
\batitle{Impact of nonequilibrium fluctuations on prethermal dynamical phase
  transitions in long-range interacting spin chains}.
\bjtitle{Phys. Rev. B}
\bvolume{99},
\bfpage{045128}
(\byear{2019})
\doiurl{10.1103/PhysRevB.99.045128}
\end{barticle}
\endbibitem

%%% 31
\bibitem[\protect\citeauthoryear{Marino et~al.}{2022}]{marino2022}
\begin{barticle}
\bauthor{\bsnm{Marino}, \binits{J.}},
\bauthor{\bsnm{Eckstein}, \binits{M.}},
\bauthor{\bsnm{Foster}, \binits{M.S.}},
\bauthor{\bsnm{Rey}, \binits{A.M.}}:
\batitle{Dynamical phase transitions in the collisionless pre-thermal states of
  isolated quantum systems: theory and experiments}.
\bjtitle{Reports on Progress in Physics}
\bvolume{85}(\bissue{11}),
\bfpage{116001}
(\byear{2022})
\doiurl{10.1088/1361-6633/ac906c}
\end{barticle}
\endbibitem

%%% 32
\bibitem[\protect\citeauthoryear{Unanyan et~al.}{2005}]{Unanyan2005}
\begin{barticle}
\bauthor{\bsnm{Unanyan}, \binits{R.G.}},
\bauthor{\bsnm{Ionescu}, \binits{C.}},
\bauthor{\bsnm{Fleischhauer}, \binits{M.}}:
\batitle{Many-particle entanglement in the gaped antiferromagnetic lipkin
  model}.
\bjtitle{Phys. Rev. A}
\bvolume{72},
\bfpage{022326}
(\byear{2005})
\doiurl{10.1103/PhysRevA.72.022326}
\end{barticle}
\endbibitem

%%% 33
\bibitem[\protect\citeauthoryear{Zunkovic et~al.}{2016}]{vzunkovivc2016}
\begin{barticle}
\bauthor{\bsnm{Zunkovic}, \binits{B.}},
\bauthor{\bsnm{Silva}, \binits{A.}},
\bauthor{\bsnm{Fabrizio}, \binits{M.}}:
\batitle{Dynamical phase transitions and loschmidt echo in the infinite-range
  xy model}.
\bjtitle{Philosophical Transactions of the Royal Society A: Mathematical,
  Physical and Engineering Sciences}
\bvolume{374}(\bissue{2069}),
\bfpage{20150160}
(\byear{2016})
\doiurl{10.1098/rsta.2015.0160}
\end{barticle}
\endbibitem

%%% 34
\bibitem[\protect\citeauthoryear{Das et~al.}{2006}]{das2006}
\begin{barticle}
\bauthor{\bsnm{Das}, \binits{A.}},
\bauthor{\bsnm{Sengupta}, \binits{K.}},
\bauthor{\bsnm{Sen}, \binits{D.}},
\bauthor{\bsnm{Chakrabarti}, \binits{B.K.}}:
\batitle{Infinite-range ising ferromagnet in a time-dependent transverse
  magnetic field: Quench and ac dynamics near the quantum critical point}.
\bjtitle{Phys. Rev. B}
\bvolume{74},
\bfpage{144423}
(\byear{2006})
\doiurl{10.1103/PhysRevB.74.144423}
\end{barticle}
\endbibitem

%%% 35
\bibitem[\protect\citeauthoryear{Salas~S et~al.}{2022}]{salas2022}
\begin{barticle}
\bauthor{\bsnm{Salas~S}, \binits{A.H.}},
\bauthor{\bsnm{Altamirano}, \binits{G.C.}},
\bauthor{\bsnm{Martinez~H}, \binits{L.J.}}:
\batitle{Analytical solution to the generalized complex duffing equation}.
\bjtitle{The Scientific World Journal}
\bvolume{2022},
\bfpage{2711466}
(\byear{2022})
\doiurl{10.1155/2022/2711466}
\end{barticle}
\endbibitem

%%% 36
\bibitem[\protect\citeauthoryear{Krech}{1997}]{krech1997}
\begin{barticle}
\bauthor{\bsnm{Krech}, \binits{M.}}:
\batitle{Casimir forces in binary liquid mixtures}.
\bjtitle{Phys. Rev. E}
\bvolume{56},
\bfpage{1642}--\blpage{1659}
(\byear{1997})
\doiurl{10.1103/PhysRevE.56.1642}
\end{barticle}
\endbibitem

%%% 37
\bibitem[\protect\citeauthoryear{Solinas et~al.}{2008}]{solinas2008}
\begin{barticle}
\bauthor{\bsnm{Solinas}, \binits{P.}},
\bauthor{\bsnm{Ribeiro}, \binits{P.}},
\bauthor{\bsnm{Mosseri}, \binits{R.}}:
\batitle{Dynamical properties across a quantum phase transition in the
  lipkin-meshkov-glick model}.
\bjtitle{Phys. Rev. A}
\bvolume{78},
\bfpage{052329}
(\byear{2008})
\doiurl{10.1103/PhysRevA.78.052329}
\end{barticle}
\endbibitem

%%% 38
\bibitem[\protect\citeauthoryear{Baksic and Ciuti}{2014}]{Baksic2014}
\begin{barticle}
\bauthor{\bsnm{Baksic}, \binits{A.}},
\bauthor{\bsnm{Ciuti}, \binits{C.}}:
\batitle{Controlling discrete and continuous symmetries in superradiant phase
  transitions with circuit qed systems}.
\bjtitle{Phys. Rev. Lett.}
\bvolume{112},
\bfpage{173601}
(\byear{2014})
\doiurl{10.1103/PhysRevLett.112.173601}
\end{barticle}
\endbibitem

%%% 39
\bibitem[\protect\citeauthoryear{Soriente et~al.}{2018}]{Soriente2018}
\begin{barticle}
\bauthor{\bsnm{Soriente}, \binits{M.}},
\bauthor{\bsnm{Donner}, \binits{T.}},
\bauthor{\bsnm{Chitra}, \binits{R.}},
\bauthor{\bsnm{Zilberberg}, \binits{O.}}:
\batitle{Dissipation-induced anomalous multicritical phenomena}.
\bjtitle{Phys. Rev. Lett.}
\bvolume{120},
\bfpage{183603}
(\byear{2018})
\doiurl{10.1103/PhysRevLett.120.183603}
\end{barticle}
\endbibitem

%%% 40
\bibitem[\protect\citeauthoryear{Li et~al.}{2018}]{BLi2019}
\begin{barticle}
\bauthor{\bsnm{Li}, \binits{B.}},
\bauthor{\bsnm{Gao}, \binits{C.}},
\bauthor{\bsnm{Xianlong}, \binits{G.}},
\bauthor{\bsnm{Wang}, \binits{P.}}:
\batitle{Critical behavior of the order parameter at the nonequilibrium phase
  transition of the ising model}.
\bjtitle{Journal of Physics: Condensed Matter}
\bvolume{31}(\bissue{7}),
\bfpage{075801}
(\byear{2018})
\doiurl{10.1088/1361-648X/aaf6cd}
\end{barticle}
\endbibitem

%%% 41
\bibitem[\protect\citeauthoryear{Huang et~al.}{2018}]{Huang2018}
\begin{barticle}
\bauthor{\bsnm{Huang}, \binits{Y.}},
\bauthor{\bsnm{Li}, \binits{T.}},
\bauthor{\bsnm{Yin}, \binits{Z.-q.}}:
\batitle{Symmetry-breaking dynamics of the finite-size lipkin-meshkov-glick
  model near ground state}.
\bjtitle{Phys. Rev. A}
\bvolume{97},
\bfpage{012115}
(\byear{2018})
\doiurl{10.1103/PhysRevA.97.012115}
\end{barticle}
\endbibitem

%%% 42
\bibitem[\protect\citeauthoryear{Stitely et~al.}{2022}]{stitely2022}
\begin{barticle}
\bauthor{\bsnm{Stitely}, \binits{K.C.}},
\bauthor{\bsnm{Giraldo}, \binits{A.}},
\bauthor{\bsnm{Krauskopf}, \binits{B.}},
\bauthor{\bsnm{Parkins}, \binits{S.}}:
\batitle{Lasing and counter-lasing phase transitions in a cavity-qed system}.
\bjtitle{Phys. Rev. Res.}
\bvolume{4},
\bfpage{023101}
(\byear{2022})
\doiurl{10.1103/PhysRevResearch.4.023101}
\end{barticle}
\endbibitem

%%% 43
\bibitem[\protect\citeauthoryear{Armitage and Eberlein}{2006}]{armitage2006}
\begin{bbook}
\bauthor{\bsnm{Armitage}, \binits{J.V.}},
\bauthor{\bsnm{Eberlein}, \binits{W.F.}}:
\bbtitle{Elliptic Functions},
pp. \bfpage{1}--\blpage{24}.
\bpublisher{Cambridge University Press},
\blocation{London}
(\byear{2006}).
\doiurl{10.1017/CBO9780511617867} .
\burl{https://doi.org/10.1017/CBO9780511617867}
\end{bbook}
\endbibitem

%%% 44
\bibitem[\protect\citeauthoryear{Wang and Guo}{1989}]{wangzhuxi1989}
\begin{bbook}
\bauthor{\bsnm{Wang}, \binits{Z.X.}},
\bauthor{\bsnm{Guo}, \binits{D.R.}}:
\bbtitle{Special Functions},
pp. \bfpage{387}--\blpage{416}.
\bpublisher{world scientific},
\blocation{Singapore}
(\byear{1989}).
\doiurl{10.1142/0653} .
\burl{https://doi.org/10.1142/0653}
\end{bbook}
\endbibitem

%%% 45
\bibitem[\protect\citeauthoryear{Castanos et~al.}{2006}]{castanos2006}
\begin{barticle}
\bauthor{\bsnm{Castanos}, \binits{O.}},
\bauthor{\bsnm{Lopez-Pena}, \binits{R.}},
\bauthor{\bsnm{Hirsch}, \binits{J.G.}},
\bauthor{\bsnm{Lopez-Moreno}, \binits{E.}}:
\batitle{Classical and quantum phase transitions in the lipkin-meshkov-glick
  model}.
\bjtitle{Phys. Rev. B}
\bvolume{74},
\bfpage{104118}
(\byear{2006})
\doiurl{10.1103/PhysRevB.74.104118}
\end{barticle}
\endbibitem

%%% 46
\bibitem[\protect\citeauthoryear{Ribeiro et~al.}{2007}]{Ribeiro2007}
\begin{barticle}
\bauthor{\bsnm{Ribeiro}, \binits{P.}},
\bauthor{\bsnm{Vidal}, \binits{J.}},
\bauthor{\bsnm{Mosseri}, \binits{R.}}:
\batitle{Thermodynamical limit of the lipkin-meshkov-glick model}.
\bjtitle{Phys. Rev. Lett.}
\bvolume{99},
\bfpage{050402}
(\byear{2007})
\doiurl{10.1103/PhysRevLett.99.050402}
\end{barticle}
\endbibitem

\end{thebibliography}

\end{document}